\documentclass[conference]{IEEEtran}

\IEEEoverridecommandlockouts

\usepackage{cite}
\usepackage{amsmath,amssymb,amsfonts}
\usepackage{algorithmic}
\usepackage{graphicx}
\usepackage{textcomp}
\usepackage{xcolor}

\usepackage{multirow}
\usepackage{paralist}
\usepackage{hyperref}
\usepackage{xspace}
\usepackage{tcolorbox}

\usepackage{tabularx,booktabs,xcolor,colortbl}
\definecolor{rulegray}{gray}{0.6}
\definecolor{headgray}{gray}{0.92}

\tcbuselibrary{listings, breakable}

\graphicspath{{./}{./fig/}}

\newcommand{\sstitle}[1]{\noindent\textbf{#1.\/}}

\newcommand{\toolname}{\texttt{FINER-SQL}\xspace}

\def\Snospace~{\S{}}

\newcommand\Mark[1]{\textsuperscript#1}
\def\ua{\Mark{*}}

\begin{document}

\setlength{\belowdisplayskip}{1pt}
\setlength{\belowdisplayshortskip}{1pt}
\setlength{\abovedisplayskip}{1pt}
\setlength{\abovedisplayshortskip}{1pt}

\title{FINER-SQL: Boosting Small Language Models for Text-to-SQL}

\author{Thanh Dat Hoang\Mark{1}, 
Thanh Trung Huynh\Mark{2},
Matthias Weidlich\Mark{3},
Thanh Tam Nguyen\Mark{1},\\
Tong Chen\Mark{4},
Hongzhi Yin\Mark{4}\ua\thanks{\ua Corresponding authors}, 
Quoc Viet Hung Nguyen\Mark{1}\ua%
\vspace{1.6mm}\\
\fontsize{9}{9}\selectfont\rmfamily\itshape
\small 
\Mark{1}Griffith University (Australia),
\Mark{2}VinUniversity (Vietnam),
\Mark{3}Humboldt-Universitat zu Berlin (Germany),
\Mark{4}The University of Queensland (Australia)
}

\maketitle

\begin{abstract}

Large language models have driven major advances in Text-to-SQL generation. However, they suffer from high computational cost, long latency, and data privacy concerns, which make them impractical for many real-world applications. A natural alternative is to use small language models (SLMs), which enable efficient and private on-premise deployment. Yet, SLMs often struggle with weak reasoning and poor instruction following. Conventional reinforcement learning methods based on sparse binary rewards (0/1) provide little learning signal when the generated SQLs are incorrect, leading to unstable or collapsed training. To overcome these issues, we propose \toolname, a scalable and reusable reinforcement learning framework that enhances SLMs through \emph{fine-grained execution feedback}. Built on group relative policy optimization, \toolname replaces sparse supervision with dense and interpretable rewards that offer continuous feedback even for incorrect SQLs. It introduces two key reward functions: a \emph{memory reward}, which aligns reasoning with verified traces for semantic stability, and an \emph{atomic reward}, which measures operation-level overlap to grant partial credit for structurally correct but incomplete SQLs. This approach transforms discrete correctness into continuous learning, enabling stable, critic-free optimization. Experiments on the BIRD and Spider benchmarks show that \toolname achieves up to 67.73\% and 85\% execution accuracy with a 3B model -- matching much larger LLMs while reducing inference latency to 5.57~s/sample. These results highlight a cost-efficient and privacy-preserving path toward high-performance Text-to-SQL generation. Our code is available at \url{https://github.com/thanhdath/finer-sql}.

\end{abstract}


\section{Introduction}
\label{sec:intro}

Text-to-SQL aims to translate natural language utterances into executable SQL queries, enabling non-experts to interact with databases seamlessly. 
The task is inherently challenging due to the need for compositional reasoning, schema linking, and handling ambiguous user intents~\cite{wang2020rat,chang2023drspider,gan-etal-2021-exploring,hung2019handling,nguyen2026handling,nguyen2024handling}. 
Recent years have seen remarkable progress driven by large language models (LLMs), which achieve state-of-the-art performance on benchmarks such as Spider and BIRD~\cite{yu-etal-2018-spider,li2024can,li2024codes}. 
Most of these methods rely on large open-source models or proprietary APIs (e.g., GPT-4, Claude), achieving high accuracy through chain-of-thought reasoning and multi-stage pipelines~\cite{talaei2024chess,pourreza2024chase,pham2026modeval,ren2022prototype,pham2024dual}.

\begin{figure}[t]
  \centering
  \includegraphics[width=1\linewidth]{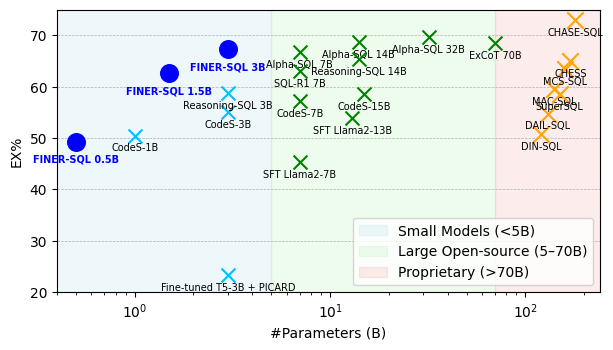}
  \vspace{-2.5em}
  \caption{Execution accuracy (EX\%) of \toolname on BIRD dev. 
  Although SLMs are often overlooked due to their 
  perceived weakness in reasoning, our 3B model outperforms many $>$14B 
  baselines and even proprietary systems.}
  \label{fig:motivating}
  \vspace{-1.5em}
\end{figure}

Despite their success, LLM-based solutions introduce significant practical constraints. 
First, relying on external APIs poses privacy risks, as sensitive database schemas and user queries must be transmitted to third-party servers, potentially violating confidentiality~\cite{abedini2025masksql,hui2025privacypad,nguyen2025privacy,nguyen2023poisoning,nguyen2023detecting}. 
Second, the cost of inference at scale can be extremely high. 
For instance, CHESS-style schema linking consumes over 340K tokens per request (over \$850 per 1{,}000 requests with GPT-4o), even before generating and selecting SQL candidates, as shown in recent analyses~\cite{chung2025long, talaei2024chess}. 
Such expenses are unsustainable for many real-world deployments, motivating the adoption of self-hosted solutions that are both cost-efficient and privacy-preserving~\cite{touvron2023llama,jiang2023mistral,huynh2023efficient,duong2022efficient,nguyen2024portable}.

A natural direction is to employ small language models (SLMs, 0.5B--3B parameters), which run efficiently on a single 12--24GB GPU (e.g., RTX 4090). 
SLMs are appealing for lightweight, on-premise deployment, but applying them to Text-to-SQL remains nontrivial. 
They suffer from (i) weak reasoning capability, leading to poor SQL generation, and (ii) limited instruction-following and formatting, producing invalid or incomplete SQL~\cite{nguyen2024manipulating,nguyen2015smart,nguyen2015tag}. 
As a result, most SQL queries from SLMs are incorrect or unexecutable, leading to very low accuracy.

Recent advances in reinforcement learning (RL) have improved the reasoning abilities of LLMs, often using policy optimization methods such as PPO~\cite{schulman2017proximal} or GRPO~\cite{shao2024deepseekmath}. These works typically train models of at least 7B parameters, using execution accuracy as the primary reward—a binary signal that gives 1 for a fully correct query and 0 otherwise~\cite{yao2025arctic, zhai2025excot,nguyen2022model,duong2022deep}. While such sparse rewards can help large models refine their reasoning, they are ineffective for SLMs, where most attempts fail and thus receive zero signal. As a result, reinforcement learning on small models often collapses or stagnates. To be effective, SLMs require fine-grained and informative rewards that can recognize partial correctness and provide continuous feedback even when the final SQL is wrong.

In this work, we introduce \textbf{\toolname}~\cite{finersql} (Fine-grained Execution feedback for Reinforcement learning in Text-to-SQL), a reinforcement learning framework tailored for small language models. Unlike prior LLM-heavy approaches that rely on 7B--32B parameter models or external APIs, \toolname targets the SLM regime ($\leq$3B), where failures dominate and dense rewards are essential. The model is trained with Group Relative Policy Optimization (GRPO), avoiding a separate value model, and augments standard execution and format rewards with two dense signals: (i) \emph{memory rewards} compare the model's reasoning to a bank of verified traces, offering stability even when execution fails; (ii) \emph{atomic rewards} score overlap between predicted and gold queries at the operation level, giving credit to partially aligned SQLs. These signals turn binary correctness into continuous feedback, revealing how far an incorrect query is from the target and which components need refinement.

As shown in \autoref{fig:motivating}, despite their compact size, our \toolname 3B variant attains 67.73\% execution accuracy on the BIRD dataset, outperforming many open-source models exceeding 14B parameters and even some proprietary systems.

We summarize our contributions as follows:
\begin{compactitem}
    \item We propose \toolname, a scalable and reusable reinforcement learning framework designed to boost Text-to-SQL performance of small models through fine-grained execution feedback.
    \item We introduce a memory reward mechanism that efficiently evaluates the reasoning traces of SLMs against verified reasoning patterns, enhancing logical consistency and training stability.
    \item We design atomic operation rewards that compute fine-grained structural similarity between generated and gold SQLs, mitigating the sparse reward problem.
    \item Comprehensive experiments on Spider and BIRD show that \toolname significantly improves execution accuracy and generalization of small models, matching much larger LLMs at a fraction of the cost. Our 3B model achieves 67.73\% execution accuracy on BIRD Dev with only 5.57s per sample.
\end{compactitem}

In the remainder, \autoref{sec:related} discusses related work.
\autoref{sec:model_problem} formulates the Text-to-SQL problem and RL challenges for SLMs. \autoref{sec:rlef} presents the \toolname framework and its two-stage design. \autoref{sec:exp} reports experimental results and conclusions (\autoref{sec:con}).

\section{Related Work}
\label{sec:related}

\sstitle{Reinforcement Learning for LLMs}
Reinforcement learning has been widely used to enhance the reasoning abilities of large language models (e.g. DeepSeek-R1~\cite{guo2025deepseek}, OpenAI’s O-series) in mathematics, logic, and code generation via reward-based optimisation. Building on the Reinforcement Learning from Human Feedback (RLHF) paradigm~\cite{bai2022training}, these methods align model behaviour with human preferences using a reward model and Proximal Policy Optimisation (PPO)\cite{schulman2017proximal,nguyen2014reconciling,nguyen2023isomorphic} under a KL constraint. Offline approaches such as Direct Preference Optimisation (DPO)\cite{rafailov2023direct} and Odds-Ratio Preference Optimisation (ORPO)\cite{hong2024orpo,huynh2024fast,zhao2021eires} eliminate explicit reward models and directly optimise preference pairs for improved stability and lower cost. In contrast, online methods including PPO and Group Relative Policy Optimisation (GRPO)\cite{shao2024deepseekmath} rely on real-time rollouts, with GRPO further removing the value model and estimating advantages via relative candidate scoring.
Our work adapts GRPO to train SLMs for Text-to-SQL, improving execution accuracy through feedback-based optimisation without a separate value model.

\sstitle{Text-to-SQL}
Text-to-SQL research has evolved from early neural models such as Seq2SQL~\cite{zhong2017seq2sql} and SQLNet~\cite{xu2017sqlnet} to grammar- and relation-aware systems including IRNet~\cite{guo2019towards}, SyntaxSQLNet~\cite{yu2018syntaxsqlnet}, and RAT-SQL~\cite{wang2020rat}, with constrained decoding methods like PICARD~\cite{scholak2021picard,thang2015evaluation,nguyen2020factcatch} improving syntactic validity. The emergence of large pretrained language models shifted the paradigm toward prompting-based and multi-stage pipelines (e.g. SQL-PaLM~\cite{sun2023sql}, DIN-SQL~\cite{pourreza2023dinsql}, CHESS~\cite{talaei2024chess,nguyen2025device,pham2025multilingual,nguyen2024multi}), integrating planning, reranking, and execution feedback but relying heavily on proprietary LLMs. More recently, reinforcement learning approaches such as Arctic-Text-to-SQL-R1~\cite{yao2025arctic}, SQL-o1~\cite{lyu2025sql}, and Ex-CoT~\cite{zhai2025excot} have shown that execution feedback improves compositional generalization, though these methods typically target large ($\geq$7B) models and depend on sparse binary rewards, limiting applicability to smaller models~\cite{thang2022nature,trung2022learning,huynh2021network}.
Our \toolname introduces dense, interpretable rewards that provide continuous feedback on reasoning and SQL structure, enabling effective reinforcement learning for small models.

\section{Model and Challenges}
\label{sec:model_problem}

\subsection{Problem Formulation}
\label{sec:problem}

Text-to-SQL addresses the task of generating an SQL query $\mathcal{Y}$ that
corresponds to a given natural language question $q$. 
$$
\mathcal{Y}=f(q, {S}, {K} \mid \boldsymbol{\theta}),
$$
where the function $f(\cdot \mid \boldsymbol{\theta})$ represents a
generative model (e.g., a neural network) with learnable parameters
$\boldsymbol{\theta}$~\cite{nguyen2018if,toan2018diversifying}.
This query
is constructed based on a database schema ${S}$ and, optionally, an
external knowledge base ${K}$. The database schema ${S}$ is
defined by
a set of tables $\left\{{T}_1, {T}_2, \ldots,
{T}_m\right\}$,
a set of columns
$\left\{{C}_1, {C}_2, \ldots, {C}_n\right\}$, and
a set of foreign key relations $\left\{{R}_1, {R}_2, \ldots, {R}_k\right\}$.
The optional external
knowledge ${K}$ provides context for the schema, aiding in
generating more accurate SQL in ambiguous situations~\cite{hung2017answer,nguyen2017argument,nguyen2023example}.

\subsection{Challenges}
\label{sec:design_principles}

We argue that any reinforcement learning approach applied to SLMs for Text-to-SQL must effectively address the following fundamental challenges:

\smallskip
\textit{(R1) Weak Reasoning Capability.}
SLM-based Text-to-SQL systems must address limited reasoning capabilities. Chain-of-thought prompting, effective for large models, is unreliable for SLMs, often yielding illogical reasoning~\cite{wei2022chain} and inaccurate SQL generation on complex schemas~\cite{li2024dawn}.

\smallskip
\textit{(R2) Low Instruction Following Capability.}
SLMs struggle with instruction following due to limited size~\cite{murthy2024evaluating} and lack of diverse tuning data like InFoBench~\cite{qin-etal-2024-infobench} and IFEval~\cite{zhou2023instruction}. Consequently, they generalize poorly to complex instructions and frequently ignore structural constraints (e.g., $<$think$>$ tags), hindering reliable SQL extraction.

\smallskip
\textit{(R3) Sparse Reward Issue.} 
RL for Text-to-SQL typically uses sparse execution rewards~\cite{zhong2017seq2sql,nguyen2025finetuning}, offering no credit for partial correctness (e.g., valid joins) and hindering fine-grained policy learning~\cite{lightman2024letsverify,yuan2024self,lu2024autopsv,shao2024deepseekmath}. Since SLMs often generate invalid initial SQLs, this results in negligible learning signals. Prior work in math and code indicates that process-level rewards improve convergence~\cite{lightman2024letsverify,wang2024math,yuan2024self}, highlighting the need for shaped, step-level feedback in SLMs.

\smallskip
\textit{(R4) High Cost of Explicit Reward Modeling.}
Process Reward Models (PRMs) offer fine-grained feedback~\cite{lightman2024letsverify,rafailov2023direct,wang2024math} but face high costs and adaptation needs~\cite{shao2024deepseekmath}. In Text-to-SQL, evaluating schema linking and partial executions adds complexity. Commercial verifiers often yield unreliable judgments~\cite{li2025alphasql,li2024can} and incur prohibitive costs (e.g., \$2400 for 10k queries~\cite{pourreza2025reasoning}), making them impractical for iterative RL.

\section{\toolname: Reinforcement Learning with Fine-Grained Execution Feedback}
\label{sec:rlef}

\begin{figure*}[!ht]
    \centering
    \includegraphics[width=0.79\linewidth]{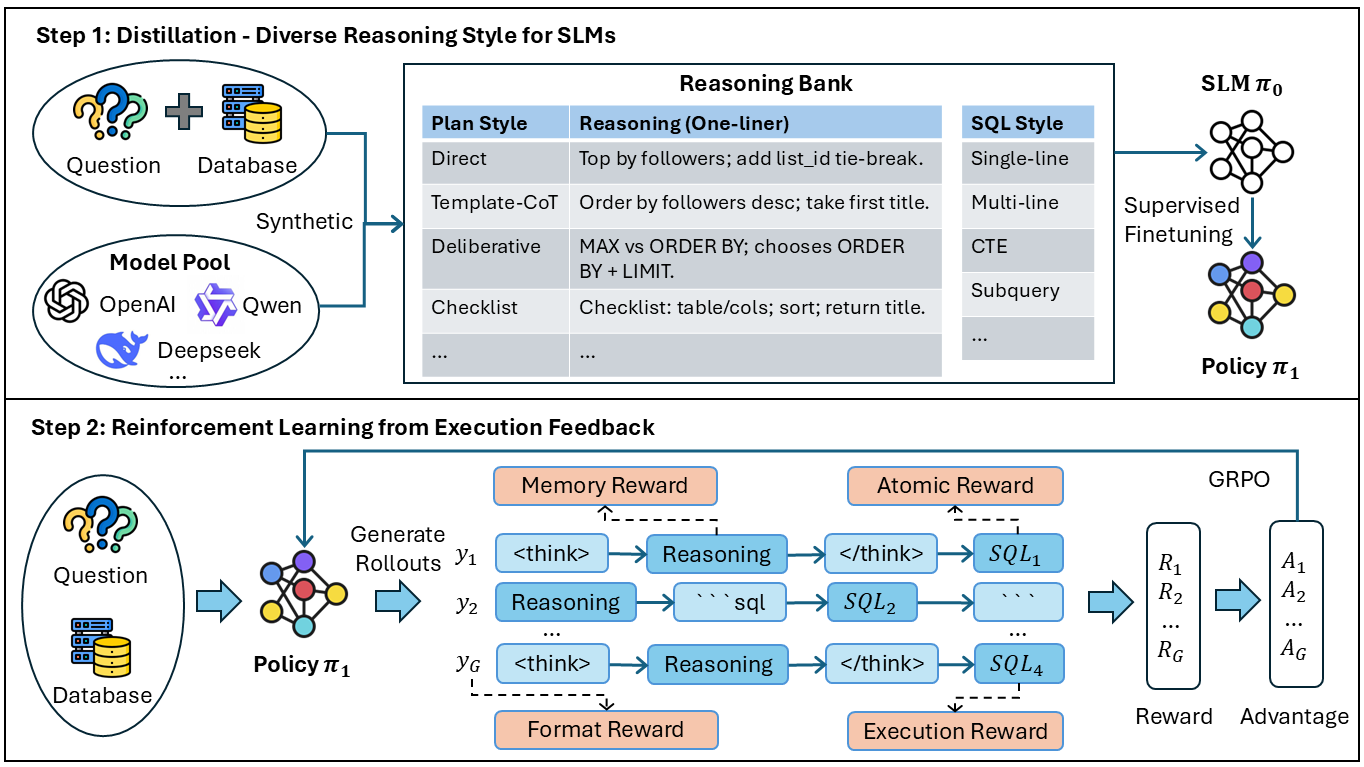}
    \vspace{-0.5em}
    \caption{Overall training pipeline of \toolname. 
    \textbf{Step~1: Distillation - Diverse Reasoning Style for SLMs.}
    Teacher models (e.g., GPT-4o, DeepSeek-R1, Qwen-2.5-72B) are prompted with question-schema pairs to produce reasoning traces and SQLs, forming a \textit{Reasoning Bank} with diverse plan and SQL styles. 
    The SLM is initialized via supervised fine-tuning on this corpus to obtain policy~$\pi_{1}$ with structured reasoning ability. 
    \textbf{Step~2: Reinforcement Learning from Execution Feedback.}
    The finetuned policy generates multiple rollouts per query and is optimized using GRPO with dense, interpretable reward signals—including \textit{format}, \textit{memory}, \textit{execution}, and \textit{atomic} rewards—to improve reasoning faithfulness and SQL accuracy.}
    \label{fig:pipeline_overview}
    \vspace{-1.5em}
\end{figure*}

Motivated by the challenges identified in \autoref{sec:design_principles}, we propose \toolname, 
a two-stage framework that enhances Text-to-SQL performance of SLMs through 
(1) reasoning distillation from large reasoning models and 
(2) reinforcement learning with fine-grained execution feedback, 
as illustrated in \autoref{fig:pipeline_overview}.

\subsection{Distillation - Diverse Reasoning Style for SLMs}
\label{sec:distillation}
\paragraph{Reasoning Bank Construction}
To address weak reasoning and limited instruction-following capabilities (R1, R2), 
we first distill diverse reasoning knowledge from a pool of large reasoning models, 
including DeepSeek-R1, GPT-oss-120B, Qwen-2.5-72B-Instruct, and OpenAI GPT-4o.  
Given a question-schema pair, each teacher model is prompted with a structured 
instruction consisting of two messages: a \textit{system prompt} that defines the 
response rules and a \textit{user prompt} that provides the database schema, 
natural-language question, and optional external knowledge.

\begin{center}
\begin{tcolorbox}[width=0.95\linewidth,title=System Prompt for Teacher Models,
left=1mm,
right=1mm,
label={lst:system_prompt}
]
\tiny
\vspace{-0.2cm}
\begin{verbatim}
You are a meticulous SQL expert. Generate a 
single, correct SQL query for the user question 
and the provided database schema.
Follow this exact response format:
<think>
Briefly reason about the steps to form the SQL.
</think>
SQL statement

Rules:
- Output exactly one SQL statement.
- The SQL must be executable on SQLite.
- Do not include any explanatory text outside the 
<think> section.
- After </think>, output one SQL statement only. 
Do not include any extra text, tags.
\end{verbatim}
\vspace{-0.2cm}
\end{tcolorbox}
\end{center}

As shown in the system prompt above, the prompt explicitly constrains teachers to output a concise reasoning trace enclosed within $<$think$>$ tags followed by a single executable SQL statement for SQLite3.  For reasoning models (e.g., DeepSeek-R1, GPT-oss-120B), the line ``Follow this exact response format:'' is omitted, since these models natively generate reasoning enclosed within $<$think$>$...$<$/think$>$ and we just need to modify the prompt so that the final SQL is returned without additional instruction. The accompanying user prompt provides the structured task inputs, namely the database schema, the natural-language question, and any optional external knowledge. This design enforces a unified output structure across all teacher models, ensuring that each completion conforms to the canonical form:
\[
\langle q, \text{schema} \rangle
\rightarrow
\langle <think>\text{reasoning}</think>\text{SQL}\rangle
\]

The resulting multi-source reasoning traces are aggregated into a 
\textit{Reasoning Bank}, providing the SLM with diverse reasoning styles and 
compositional strategies across different teacher architectures.  
We then perform supervised fine-tuning (SFT) on this distilled corpus to 
transfer structured reasoning and SQL generation ability to the smaller model.  
This step also enables the SLM to quickly adapt to the required output format, 
as it learns to generate reasoning traces followed by valid SQL structures from 
the beginning of training.  
Importantly, we do not filter out teacher samples with failed executions, 
since the purpose of this stage is not to achieve execution correctness but to 
expose the model to a wide variety of reasoning trajectories and compositional 
patterns.  
Although the execution accuracy of teacher models on the BIRD training set is 
around 50--60\%, these imperfect yet diverse examples still provide valuable 
reasoning supervision that helps the SLM develop robust structural understanding.  
We generated 37.6K distillation samples (4 teachers $\times$ BIRD training set) for only $\sim$\$30. This cost was minimized by employing GRAST-SQL \cite{hoang2025scaling}, an efficient schema-linking method, to restrict inputs to the top-30 columns instead of the entire large schema.

\paragraph{Supervised Fine-Tuning Process}
Given a Reasoning Bank containing diverse reasoning-SQL pairs distilled from multiple teacher models, we perform supervised fine-tuning to transfer their structured reasoning and compositional knowledge to the smaller model. During fine-tuning, we optimize the SLM only on the completion tokens -- i.e., the reasoning and SQL segments -- while excluding prompt tokens such as schema descriptions or external context. This ensures that the model learns to reproduce reasoning paths and final SQLs without memorizing database-specific content.  
The supervised fine-tuning loss is formulated as:
\begin{equation}
\label{eq:l_completion}
    \mathcal{L}_{\text{completion}}
    = -\sum_{j=C+1}^{\tau} 
      \log P_{\theta}(y_j \mid y_{<j}, \chi).
\end{equation}
where $C$ marks the end of the prompt tokens, $\tau$ is the total sequence length, $y_t$ is the target completion token, $\chi$ denotes the prompt, and $P_{\theta}$ is the model's predicted probability distribution parameterized by $\theta$.  

Through this process, the SLM acquires structured reasoning ability, format obedience, 
and generalization over diverse reasoning patterns—forming a strong initialization 
for the subsequent reinforcement learning stage.

\subsection{Reinforcement Learning from Execution Feedback}
\label{sec:grpo_training}

To address the challenges of sparse and costly supervision signals (R3, R4), the distilled SLM is further optimized through reinforcement learning using fine-grained execution feedback.  
For each training instance, the model samples a group of $G$ rollouts $\{r_1,\dots,r_G\}$, where each rollout includes a reasoning trace and its corresponding SQL prediction.  
Instead of relying on a binary execution signal, we design a \textit{dense composite reward} that reflects different dimensions of SQL quality and reasoning consistency.  
The reward comprises four complementary components:

\begin{compactitem}
    \item \textbf{Format Reward} encourages the model to follow the standardized $<$think$>$-SQL structure, where the reasoning process and executable SQL are clearly separated for reliable parsing and evaluation.  
Although the distilled policy $\pi_{1}$ from \autoref{sec:distillation} often generates correctly formatted outputs, the format reward remains essential during reinforcement learning, as exploration may cause the model to deviate from the expected structure.  
Formally,
\[
R_{\text{format}}(s) =
\begin{cases}
1, & \text{if the output follows the format}\\[4pt]
0, & \text{otherwise.}
\end{cases}
\]
    \item \textbf{Execution Reward} assesses the executable correctness of the generated SQL.  
Queries that fail to run receive zero reward.  
If the SQL executes successfully but produces an incorrect result, a base credit of $1.0$ is assigned for syntactic and runtime validity.  
When the predicted and ground-truth outputs match exactly, the model receives the maximum reward of $2.0$:  
\[
R_{\text{exec}}(s \mid q) =
\begin{cases}
0, & \text{if the SQL fails to execute;}\\
1, & \text{if executable without syntax error;}\\
2, & \text{if results match exactly.}
\end{cases}
\]

    \item \textbf{Memory Reward} evaluates the semantic alignment of a generated reasoning trace with a set of verified reasoning prototypes, particularly when the SQL execution fails.
It assigns a continuous score $R_{\text{mem}}\!\in[0,1]$ based on embedding similarity between the generated reasoning and gold reasoning traces retrieved from diverse databases.  
This reward serves two purposes: (i) it estimates the likelihood that the reasoning path is logically valid, and (ii) it regularizes the model to follow the diverse reasoning styles distilled from teacher models rather than generating noisy or unfocused reasoning.  
A detailed formulation is provided in \autoref{sec:memory_reward}.

\item \textbf{Atomic Reward} captures structural similarity between the predicted and reference SQLs by decomposing each query into a list of atomic operations—such as \texttt{FROM}, \texttt{JOIN}, and \texttt{WHERE\_PRED}—and measuring their overlap.  
Each SQL is parsed into an operation set, and the reward assigns a continuous score $R_{\text{atomic}}\!\in[0,1]$ proportional to the intersection ratio between the predicted and ground-truth operation sets.  
This fine-grained feedback provides partial credit for structurally correct components, encouraging the model to progressively assemble a complete and executable query, as detailed in~\autoref{sec:atomic_reward}. 
The atomic reward is activated only when the predicted SQL is incorrect.

\end{compactitem}

The overall reward for each rollout is defined as the unweighted sum of all components:
\[
R = R_{\text{format}} + R_{\text{exec}} + R_{\text{atomic}} + R_{\text{mem}},
\]
When an SQL fails to execute, the \textit{atomic} and \textit{memory} rewards provide fallback supervision, allowing the policy to continue learning from structurally or semantically meaningful rollouts.

For optimization, we employ GRPO~\cite{shao2024deepseekmath}, which estimates relative advantages across a group of $G$ rollouts for each instance. 
This approach eliminates the need for a separate critic or reward model, enabling stable policy improvement with reduced computational cost.

\begin{figure}[!t]
    \centering
    \includegraphics[width=0.8\linewidth]{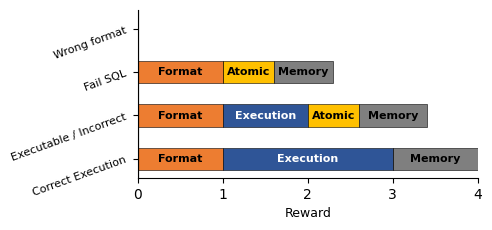}
    \vspace{-1.5em}
    \caption{
    \textbf{Reward scaling across prediction quality levels.}
    Each stacked bar shows how individual rewards (Format, Execution, Atomic, Memory) 
    contribute under increasing SQL correctness—from wrong format to fully correct execution.  
    The scaling illustrates that \toolname\ replaces binary rewards with a smooth, continuous feedback space, 
    ensuring dense and interpretable credit assignment throughout training.}
    \label{fig:reward_scaling}
    \vspace{-1.5em}
\end{figure}

\vspace{3pt}
\noindent\textbf{Reward composition and scaling.}  
To illustrate the behavior and relative range of each reward, 
\autoref{fig:reward_scaling} visualizes how rewards accumulate across representative 
prediction categories—ranging from wrong format to correct execution.  
The Format Reward (orange) is always computed first and bounded in $\{0,1\}$, 
acting as a gate that ensures structural validity.  
Subsequent rewards, Execution (blue), Atomic (yellow), and Memory (gray), expand upon this baseline, 
demonstrating how partial correctness is progressively credited even before perfect execution is achieved.  
This scaling visualization highlights the dense nature of our reward design: every stage of improvement, 
from valid syntax to full correctness, produces a measurable and interpretable reward increase.

\subsection{Memory Reward: Reasoning-SQL Semantic Alignment}
\label{sec:memory_reward}

\paragraph{Motivation}
The output of SLMs consists of a $<$think$>$ reasoning trace followed by the generated SQL, where the SQL quality directly depends on the reasoning.  
When the SQL fails to execute, evaluating only the final output discards meaningful reasoning progress and destabilizes learning.  
We therefore propose the \textit{Memory Reward}, which evaluates the reasoning itself based on its semantic similarity to successful reasoning traces stored in memory.  
This reward serves two purposes:  
(1) it promotes \emph{reasoning consistency} by anchoring updates around coherent and previously verified reasoning patterns, mitigating logical variance among the $K$ sampled candidates to prevent fragmented execution groups, thereby ensuring higher consensus during majority voting; and
(2) it acts as a \emph{semantic regularizer} that concentrates the model's probability mass on valid structural prototypes, preventing the candidate pool from being diluted by degenerate traces (e.g., excessively short sequences, uninformative content, or malformed XML tags) and directly improving the robustness of the final aggregated prediction.

\paragraph{Theoretical Formulation}

We formulate the Text-to-SQL generation process as a \textit{Goal-Augmented Partially Observable Markov Decision Process (POMDP)}, defined by the tuple
$\mathcal{M} = (\mathcal{S}, A, \mathcal{P}, \mathcal{G}, R, O)$.
The components of this process are defined as follows:
\begin{compactitem}
    \item $\mathcal{S}$ is the latent state space.
    \item $A \subset V^L$ represents the action space, where each action is a sequence sampled from the language model's vocabulary $V$, with $L$ denoting the maximum length of the generated text.
    \item $\mathcal{P}: \mathcal{S} \times A \rightarrow \mathcal{S}$ is the transition function updating the state based on the generated action.
    \item $\mathcal{G} \subset V^N$ denotes the goal space, representing the set of executable SQL queries that satisfy the user intent.
    \item $R: \mathcal{S} \times A \times \mathcal{G} \rightarrow \mathbb{R}$ represents the goal-conditioned reward function.
    \item $O$ is the set of observations $o \in O$, capturing the visible context (question, schema) available to the model.
\end{compactitem}

In this framework, the objective of the policy $\pi_\theta$ is to maximize the expected return, i.e., the expected total reward accumulated over a rollout:
\[
J(\theta)
= \mathbb{E}_{\boldsymbol{\xi} \sim \pi_\theta}
\Bigg[\sum_{t=1}^{H} R(s_t, a_t, g)\Bigg],
\]
where $\boldsymbol{\xi} = (s_1,a_1,\dots,s_H,a_H)$ denotes a rollout sampled from $\pi_\theta$,
and $\sum_{t=1}^{H} R(s_t, a_t, g)$ is the cumulative reward of that rollout.
A fundamental challenge in applying reinforcement learning to LLMs is the vastness of the action space $A \subset V^L$.
The set of valid reasoning paths $\mathcal{T}_{valid} \subset A$ leading to a goal $g \in \mathcal{G}$ is combinatorially large.
A policy trained solely on sparse execution signals ($R_{exec} \in \{0, 1\}$) tends to explore this space unconstrained, developing high variance.
Concretely, given an observation $o\in O$ (i.e., a question--schema prompt), the policy samples a group of rollouts $\{\boldsymbol{\xi}_j\}_{j=1}^{K}$ by stochastic decoding with temperature $T$, and each rollout is scored by the composite reward. 
At inference, we follow the same sampling protocol, drawing $K$ candidate generations under temperature $T$ and select the best SQL via majority voting; in practice, diverse reasoning traces often lead to diverse SQL structures, which fragment candidates into more execution groups (i.e., $K$ distinct SQLs) and thus cause vote dilution, where no single group secures a decisive majority.

To mitigate this effect, we introduce the Memory Reward ($R_{\text{mem}}$) as a regularization signal that promotes reasoning consistency: by assigning higher scores to rollouts whose reasoning traces align with verified memory prototypes, the policy shifts probability mass toward a smaller set of coherent reasoning patterns, causing candidates to cluster into fewer execution groups and yielding higher consensus during inference.
We define a manifold of verified reasoning prototypes $\mathcal{T}_{mem} \subset \mathcal{T}_{valid}$, representing the high-quality traces stored in memory during the distillation phase. 
The optimization objective is formulated as:
\[
\max_\theta \;
\mathbb{E}_{\boldsymbol{\xi} \sim \pi_\theta}
\big[ R_{exec}(\boldsymbol{\xi}) + \lambda R_{mem}(\boldsymbol{\xi}) \big].
\]
Mathematically, $R_{mem}$ reshapes the policy's energy landscape by concentrating probability mass on the verified prototypes in $\mathcal{T}_{mem}$, thereby constraining the vast search space $\mathcal{T}_{valid}$.
This significantly reduces policy variance
($\text{Var}(\pi_{\text{mem}}) \ll \text{Var}(\pi_{\text{sparse}})$),
causing sampled candidates to cluster into fewer execution groups and ensuring robust consensus for majority voting.
Notably, we assume that the reasoning patterns distilled from our selected teacher models are already high-quality; therefore, $R_{\text{mem}}$ is designed to encourage the SLM to stay close to these verified prototypes rather than exploring entirely new reasoning patterns.

\begin{figure}[!t]
    \centering
    \includegraphics[width=0.7\linewidth]{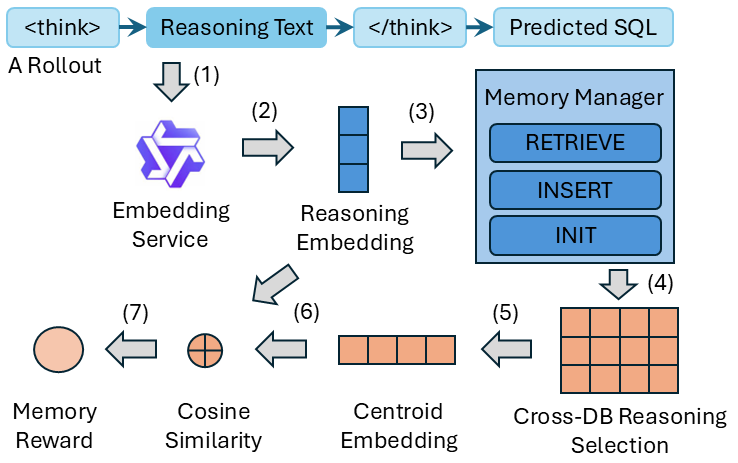}
    \vspace{-1em}
    \caption{The flow of memory reward computation and management.  
    During rollout, the reasoning text is embedded and compared with the centroid of top-$k$ similar reasoning retrieved from different databases.  
    The memory manager executes three key functions—INIT, RETRIEVE, and INSERT—to maintain a diverse set of verified reasoning traces for GRPO optimization.}
    \label{fig:memory_reward_flow}
    \vspace{-2em}
\end{figure}

\paragraph{Memory construction and management}
We define a persistent reasoning memory $\mathcal{M}$ implemented as a vector database using ChromaDB \cite{han2023comprehensive}, supporting efficient embedding retrieval and incremental updates of reasoning traces.  
Each memory entry stores a reasoning embedding, its associated metadata (sample ID, reasoning embedding, and database ID), and is managed by three operators INIT, RETRIEVE, and INSERT, as illustrated in \autoref{fig:memory_reward_flow}.

\begin{compactitem}
    \item \textbf{\texttt{INIT}: Memory Initialization.}  
    The memory $\mathcal{M}$ is initialized with high-quality reasoning traces from the teacher model DeepSeek-R1, which often produces diverse reasoning styles for distillation in \autoref{sec:distillation}. We first filter samples whose generated SQLs execute correctly, then extract their reasoning content from the $<$think$>$ blocks. Each reasoning trace is embedded into a shared semantic space using Qwen3-Embedding-0.6B~\cite{zhang2025qwen3}, and stored in ChromaDB with its metadata. This initialization builds a foundational pool of verified reasoning paths.

    \item \textbf{\texttt{RETRIEVE}: Cross-Database Semantic Retrieval.}  
    During training, for a given reasoning trace $t$ from database $d$, the RETRIEVE operator searches for top-$k$ semantically similar traces $\{r_i\}_{i=1}^k$ from other databases ($db\_id \neq d$). 
By calculating the reward based on the centroid of these $k$ embeddings, the system forces the model to align with the average structural pattern rather than overfitting to specific schema tokens. Furthermore, this design does not encourage generic hallucinations, because if the model generates hallucinated schema elements from other databases, this easily leads to a SQL syntax error on the current database, triggering a zero execution reward and penalizing the deviation.
    
\item \textbf{\texttt{INSERT}: Incremental Memory Update.}  
When a reasoning-SQL pair executes successfully, its reasoning trace is embedded and conditionally inserted into $\mathcal{M}$ using INSERT.  
Before insertion, the system applies a quality gate to filter degenerate traces based on three metrics derived from the trace text: (1) Length Heuristics reject traces that are too short ($<30$ tokens) or too long ($>2000$ tokens); (2) Informativeness ensures schema grounding by requiring a schema density (calculated as the ratio of schema column mentions to total tokens) $\ge 0.05$ with at least 2 distinct column mentions; and (3) Lexical Diversity prevents repetitive loops by requiring that at least 60\% of consecutive word pairs are unique. Valid traces then undergo similarity deduplication: the system skips addition if the cosine similarity with the top-1 existing entry is $\geq 0.9$, thereby avoiding redundancy and encouraging exploration of new reasoning paths.
Over time, $\mathcal{M}$ evolves dynamically, continually enriched by successful reasoning produced by the student model.

\end{compactitem}

These operators work together to keep the reasoning memory $\mathcal{M}$ clean and diverse. The system only stores reasoning traces that are both correct in execution and sufficiently different from what already exists in memory. If a new reasoning is too similar to previous ones, it is skipped. This mechanism helps $\mathcal{M}$ grow into a well-balanced repository that contains only effective and varied reasoning paths, guiding the model toward better generalization.

\paragraph{Reward formulation}
Let $\psi(\cdot)$ be the fixed encoder for embedding reasoning paths.
Given a question/sample $q$ (with associated database $\operatorname{db}(q)$) and the reasoning trace $t$ generated by the SLM,
we retrieve the top-$k$ verified reasoning paths from other databases
\[
\mathcal{R}(t; q)=\{\,r_i \mid \texttt{db\_id}(r_i)\neq \operatorname{db}(q)\,\}_{i=1}^k,
\]
and compute their embeddings $E=\{\psi(r_i)\}_{i=1}^k$.
We then form the centroid embedding
\[
c(t;q) = \frac{1}{k}\sum_{i=1}^{k} \psi(r_i),
\]
and normalize the vectors $\hat a=\psi(t)/\|\psi(t)\|_2$ and $\hat c=c(t;q)/\|c(t;q)\|_2$.
\[
R_{\text{mem}}(t\mid q) =
\begin{cases}
0, & \text{if the output format is invalid;}\\
\cos(\hat a, \hat c) , & \text{if the SQL executes incorrectly;}\\
1, & \text{if the SQL executes correctly.}
\end{cases}
\]
This formulation treats the reward as the direct cosine similarity between the current reasoning and the centroid of retrieved successful traces, providing a dense, interpretable signal of reasoning alignment.  
When execution succeeds, the reward saturates at $1.0$ and the reasoning trace is stored in the memory bank $\mathcal{M}$ via the \texttt{INSERT} operator.  
This design preserves stable gradients, avoids redundant scaling, and ensures that even partially correct reasoning receives proportional credit.

\paragraph{Example}
\emph{Schema (simplified):} Employee(EmpID, Name, DeptID, Salary), Department(DeptID, DeptName).  
\emph{Question:} “List employees who earn more than the average salary of their own department.”

\begin{tcolorbox}[colback=gray!20, colframe=gray!90, boxrule=0.3pt, arc=1pt, left=3pt, right=3pt, top=2pt, bottom=2pt, breakable]
\scriptsize
$<$think$> $ \\
1) Compute the average salary per department.\\
2) Compare each employee's salary with the department average.\\  
3) Return those exceeding it.\\  
$<$/think$>$\\[-2pt]
SQL: (execution fails)
\end{tcolorbox}

Even though the SQL fails, the reasoning follows a coherent logical pattern—group, aggregate, align, and compare—that aligns with verified traces retrieved from other databases (e.g., comparing students above class average or products above category mean).  
The high semantic similarity produces a strong memory reward, guiding the model to refine rather than discard this reasoning path.  

In contrast, when the model produces invalid reasoning patterns, various failure cases can occur, such as empty reasoning traces, hallucinated answers instead of SQL queries, or the generation of XLM/HTML tags without meaningful information.
These reasoning traces are semantically distant from the reasoning patterns stored in memory, leading to very low similarity scores and thus minimal memory rewards.
Such penalties discourage unstructured or hallucinated reasoning behaviors and steer the model back toward compositional, SQL-oriented reasoning that can yield executable queries.

\paragraph{Efficiency and stability}
The memory module is designed for efficiency and scalability.
Each reasoning trace is embedded once and indexed in a vector database, enabling sub-millisecond approximate nearest-neighbor retrieval even at large scale \cite{han2023comprehensive,johnson2019billion}.
The vector database not only accelerates retrieval but also ensures consistent and reliable access to reasoning traces during training.
Since the reward depends only on embedding similarity rather than SQL execution, it incurs negligible overhead and can be computed in parallel with rollouts.
This lightweight mechanism provides dense, cost-effective feedback that stabilizes GRPO optimization and improves reasoning transfer across databases.

\subsection{Atomic Reward: Operation-Level Structural Feedback}
\label{sec:atomic_reward}

Execution-based rewards assess only final correctness and fail to capture structural progress for partially correct or non-executable SQL. To address this, we introduce the \textit{Atomic Reward}, which decomposes each SQL query into \emph{atomic operations} (e.g., selections, joins, predicates) and measures overlap with the reference SQL at this level, providing continuous, interpretable feedback that reflects compositional reasoning.

\sstitle{Motivation}
A Text-to-SQL model may generate SQLs that are semantically close to the ground truth—using correct tables, joins, or filters—but differ in specific clauses.  
For instance, replacing \texttt{AVG(Salary)} with \texttt{COUNT(*)} or omitting a \texttt{GROUP BY} clause.  
While token-based or n-gram similarity metrics can capture surface-level overlap, they remain sensitive to textual styling and fail to reflect true structural alignment between SQL components.  
Execution-based rewards likewise overlook these fine-grained distinctions.  
By analyzing SQLs through atomic operations, our comparison becomes invariant to formatting and reordering, interpretable in terms of compositional structure,  
and capable of assigning partial credit to logically aligned fragments—encouraging the model to gradually assemble full correctness.

\sstitle{Atomic decomposition}
Each SQL query $s$ is parsed into \textit{atomic operations} following \autoref{tab:atomic_ops} by constructing an \textit{abstract syntax tree (AST)} and extracting clause-level primitives. Each AST node represents a distinct SQL construct (e.g., projections, predicates, aggregations, joins), enabling semantically meaningful operations independent of token order or formatting. This abstraction yields comparable atomic representations for syntactically different but equivalent SQLs; for deeply nested subqueries, operations are flattened into a single list to ensure efficient and consistent comparison.

\begin{table}[t]
    \centering
\setlength{\tabcolsep}{2pt}
\renewcommand{\arraystretch}{0.95}
\scriptsize
\caption{All defined atomic operations; each represents a meaningful SQL construct for structural comparison.}
\label{tab:atomic_ops}
\vspace{-0.5em}
\resizebox{0.9\columnwidth}{!}{
\begin{tabularx}{\columnwidth}{|l|X|}
\hline
\rowcolor{headgray}
\textbf{Atomic op} & \textbf{Meaning / Example} \\
\hline

\rowcolor{headgray}\multicolumn{2}{|l|}{\textbf{FROM / JOIN}}\\ \hline
FROM(tbl [AS a]) & Base relation; FROM(cust AS T1) \\ 
JOIN(tbl [AS a], type) & Join relation; JOIN(orders, INNER) \\ 
ON\_EQ(a.col,b.col) & Equality key; ON\_EQ(T2.id,T1.id) \\ 
ON\_PRED(op,lhs,rhs) & Non-equi join; ON\_PRED($>$,T2.qty,VALUE(0)) \\ \hline

\rowcolor{headgray}\multicolumn{2}{|l|}{\textbf{SELECT projections}}\\ \hline
SELECT\_COL(a.col) & Column; SELECT\_COL(name) \\ 
SELECT\_AGG(agg,a.col) & Aggregation; SELECT\_AGG(SUM,total) \\ 
SELECT\_EXPR(expr) & Scalar expr; SELECT\_EXPR(LOWER(name)) \\ 
DISTINCT & Distinct projection \\ \hline

\rowcolor{headgray}\multicolumn{2}{|l|}{\textbf{WHERE / HAVING}}\\ \hline
WHERE\_PRED(op,lhs,rhs) & Predicate; WHERE\_PRED(=,ct,VALUE(`AU')) \\ 
HAVING\_PRED(op,lhs,rhs) & Post-group filter \\ \hline

\rowcolor{headgray}\multicolumn{2}{|l|}{\textbf{Literals / Group / Order / Limit}}\\ \hline
VALUE(v) & Literal; VALUE(DATE `2024-01-01') \\ 
GROUP\_BY(a.col) & Group key; GROUP\_BY(name) \\ 
ORDER\_BY(expr,dir) & Ordering; ORDER\_BY(rev,DESC) \\ 
LIMIT(n) & Row cap; LIMIT(50) \\ \hline

\rowcolor{headgray}\multicolumn{2}{|l|}{\textbf{Set ops / CTEs}}\\ \hline
UNION / INTERSECT / EXCEPT & Set operation \\ 
WITH\_CTE(name) & Declare CTE; WITH\_CTE(top\_cust) \\ \hline

\rowcolor{headgray}\multicolumn{2}{|l|}{\textbf{Nested subqueries}}\\ \hline
ENTER\_SUBQUERY(role) & Enter subquery; \\
&                       ENTER\_SUBQUERY(WHERE\_SCALAR) \\ 
EXIT\_SUBQUERY & Exit subquery scope \\ 
SUBQ\_LAST & Refer to last subquery; \\
&           WHERE\_PRED(=,x,SUBQ\_LAST) \\ \hline

\rowcolor{headgray}\multicolumn{2}{|l|}{\textbf{Window / Alias}}\\ \hline
WINDOW(part,order,frame) & Window spec \\ 
SELECT\_WIN(func,args) & Windowed select \\ 
ALIAS(kind,from,to) & Rename; ALIAS(COLUMN,SUM(total),rev) \\ \hline
\end{tabularx}
}
\vspace{-2em}
\end{table}

\begin{figure}[!t]
    \centering
    \includegraphics[width=0.95\linewidth]{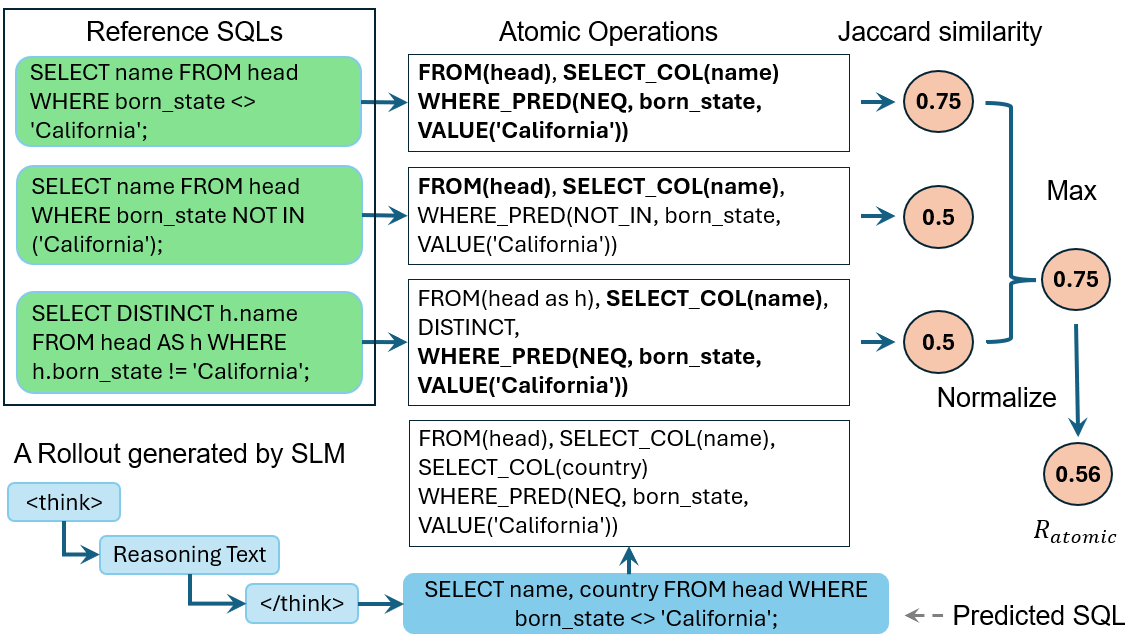}
    \vspace{-0.8em}
    \caption{
    \textbf{Atomic Reward computation pipeline.}  
    The predicted SQL is compared against a set of reference SQLs that represent equivalent but stylistically different realizations of the same intent.  
    Each comparison yields a Jaccard similarity  across atomic operations, and the final reward is the maximum among these scores.}
    \label{fig:atomic_reward_flow}
    \vspace{-2em}
\end{figure}

\sstitle{Reward computation}
Given a predicted SQL $s$ generated from a rollout,  
we parse it into a set of atomic operations $\mathcal{A}(s)$ using our Atomic Decomposition module (see the left part of \autoref{fig:atomic_reward_flow}).
Each atomic unit captures one structural action such as \texttt{SELECT\_COL}, \texttt{FROM}, or \texttt{WHERE\_PRED}, enabling direct comparison between queries at the operation level.

For each reference SQL $g_i$ in the ground-truth SQLs set $\{g_1, \ldots, g_{N'}\}$ (see the reference SQL enrichment process below), we compute a Jaccard similarity between their atomic-operation sets:
\[
J(s,g_i) \;=\;
\begin{cases}
\dfrac{|\mathcal{A}(s)\cap \mathcal{A}(g_i)|}{|\mathcal{A}(s)\cup \mathcal{A}(g_i)|}, & \text{if } |\mathcal{A}(s)\cup \mathcal{A}(g_i)|>0,\\[6pt]
0, & \text{otherwise.}
\end{cases}
\]
This metric captures the proportion of shared operations (e.g., same \texttt{SELECT\_COL(name)}, \texttt{WHERE\_PRED(=,\dots)}), normalized by the total number of distinct operations across both queries.  
As illustrated in the \emph{Atomic Operations Comparison} part of \autoref{fig:atomic_reward_flow}, 
each operation in the predicted SQL is aligned with its counterpart in the reference SQLs, 
where bold elements denote overlapping atomic operations that contribute to the overall Jaccard score.

To prevent the atomic score from saturating near~1 for high but imperfect overlaps, we apply a monotone shaping function:
\[
\phi(x) = \lambda\,x + (1-\lambda)\,\beta\,x^{\gamma},
\quad \text{with } \lambda{=}0.05,\; \beta{=}0.79,\; \gamma{=}0.20.
\]
This compresses the top end of the scale (e.g., $\phi(0.95){<}0.95$), so ``almost-right'' structures receive slightly less credit than totally matches.
The goal is to avoid over-rewarding queries that are structurally very similar yet not fully correct, while still providing smooth credit for partial matches. 
A fully correct SQL will still be reward 2.0 by the execution reward.

We first compute the maximum Jaccard similarity over all reference SQLs:
\[
J_{\max}(s) = \max_{i \in [1, N']} J(s, g_i),
\]
and then apply the shaping function to normalize the reward:
\[
R_{\text{atomic}}(s) = \phi(J_{\max}(s)).
\]
This formulation enhances robustness to structural variations by selecting the most compatible reference query among all semantically equivalent candidates.  
As a result, the reward remains stable even when the predicted SQL differs in syntax, such as through alias renaming, operator choice, or subquery restructuring, yet conveys the same semantics.

\sstitle{Reference SQL enrichment}
To increase the diversity and coverage of structurally correct SQL references, we enrich the ground-truth set using our largest distilled model (3B) in \autoref{sec:distillation}. 
For each training question, we sample $N{=}30$ candidate SQLs from the distilled model with temperature $T{=}1.0$, using the \emph{exact same prompt format} as in its SFT stage. 
This is important: since the distilled model has been explicitly trained to reproduce the correct SQL for these prompts, its generations are strongly anchored to the gold logical structure and typically remain highly similar to the ground-truth query. 
The observed diversity when decoding multiple candidates at $T{=}1.0$ mainly stems from the distillation setup itself: during distillation, a single prompt is paired with multiple teacher-produced completions (i.e., different reasoning traces and SQL) coming from diverse teachers. 
As a result, sampling $N{=}30$ completions naturally yields multiple stylistic variants in the \emph{thinking--SQL} output (e.g., aliasing, join ordering, or equivalent predicate phrasing) while preserving the underlying semantics required to answer the question. 
We then execute each candidate on the corresponding database and retain only those whose denotation matches the gold query, adding them to the reference pool $\{g_1,\ldots,g_{N'}\}$ \emph{together with the original gold SQL}.

A potential concern is \emph{spurious correctness} under empty execution result: if the gold SQL yields an empty result set, a degenerate ``hack'' query such as \texttt{WHERE 1=0} could also return the same empty output and might be mistakenly added to the reference pool. 
In our setting, however, such hacks are unlikely to appear unless the training instance itself is mislabeled, because the distilled model is trained with the same prompts to reproduce the gold SQL and thus tends to generate queries that remain structurally close to the ground truth rather than collapsing to vacuous predicates. 
Empirically, the risk of reinforcing spurious structures is minimal: on the BIRD training set, $670/9{,}428$ samples (7.1\%) produce empty gold results, yet the execution-verified candidates we retain still exhibit high structural overlap with the corresponding gold SQL.
Additionally, in a manual audit of 100 additional non-empty cases, we did not observe hack-like patterns; the main discrepancy was \texttt{SELECT} versus \texttt{SELECT DISTINCT}, which is consistent with known labeling noise in NL2SQL training data \cite{liu2025nl2sql}. 
Finally, since the atomic reward is computed as the maximum similarity over the reference pool (which always includes the gold SQL), enrichment primarily reduces false mismatches due to syntactic variations without encouraging semantically incorrect SQL.

\section{Empirical Evaluation}
\label{sec:exp}

In this section, we conduct experiments with the aim of answering the
following research questions:

\begin{compactitem}
\item[(RQ1)] Does \toolname improve execution accuracy of small language models over existing baselines? (\autoref{sec:end2end})
\item[(RQ2)] How efficient is \toolname in memory usage and inference time compared with existing methods? (\autoref{sec:performance})

\item[(RQ3)] How sensitive is model performance to hyperparameters such as temperature and candidate number? (\autoref{sec:sensitivity})
\item[(RQ4)] How does reinforcement learning with dense rewards improve SQL accuracy over training steps? (\autoref{sec:alignment_vs_sft})

\item[(RQ5)] How do different reward components (atomic, memory) contribute to the overall model performance? (\autoref{sec:ablation})
\item[(RQ6)] How well does \toolname handle SQL queries of varying complexity levels?  (\autoref{sec:complexity})
\item[(RQ7)] How does the reasoning evolve after reinforcement learning training? (\autoref{sec:reasoning_length})

\end{compactitem}

\subsection{Experimental Setup}

\sstitle{Database} We rely on two popular Text-to-SQL benchmarks:
Spider~\cite{yu-etal-2018-spider} and BIRD~\cite{li2024can,ren2024comprehensive,nguyen2026review,pham2025extensible}.
\textit{Spider} is a popular benchmark for NL2SQL translation, consisting
of 200 databases with multiple tables that cover 138 diverse domains. Spider
contains 7000 samples in a training set, a development set with 1024
samples, and a test set.
\textit{BIRD} contains 95 databases, cumulatively accounting for 33.4GB
across 37 professional domains. BIRD contains 9428 samples in
training set, a development set with 1534 samples and a hidden test set.
BIRD is more challenging, with each of BIRD's databases containing around
549K rows on average, compared to Spider's limited capacity of just 2k
rows. Also, BIRD offers evidence for a specific sample to facilitate
the generation of the right SQL query.

\sstitle{Evaluation Metrics}
We evaluate model performance using \emph{Execution Accuracy} (EX) and \emph{Pass@K}~\cite{huynh2025certified,yang2024pdc,sakong2024higher}. EX measures whether the predicted SQL produces the same execution result as the ground truth, serving as the primary indicator of correctness. Pass@K captures the likelihood that at least one of the top-$K$ generated SQL candidates executes correctly, which is particularly important for selection-based Text-to-SQL systems (e.g. CHASE-SQL \cite{pourreza2024chase}, Reasoning-SQL \cite{pourreza2025reasoning}), as a correct query among top-$K$ candidates can potentially be identified by an effective selection or ranking algorithm.

\sstitle{Baselines}
We compare \toolname against a wide range of Text-to-SQL systems, including closed-source API-based methods and open-source fine-tuned models. Although recent approaches such as OmniSQL~\cite{li2025omnisql} and Arctic-Text-to-SQL-R1~\cite{yao2025arctic} leverage large-scale synthetic corpora (e.g., SynSQL-2.5M~\cite{li2025omnisql}, Gretel-Synth~\cite{gretel-synthetic-text-to-sql-2024}), these introduce new schemas and question formulations that differ substantially from Spider and BIRD. To ensure fair, controlled comparisons that isolate training algorithms rather than data volume, we evaluate only methods trained on Spider and BIRD databases.

\textit{Closed-source API-based Methods.}
DIN-SQL~\cite{pourreza2023din} decomposes complex questions into schema linking, query classification, and self-correction before SQL synthesis. DAIL-SQL~\cite{gao2023text} combines prompt engineering with lightweight fine-tuning, selecting demonstrations by question-query similarity. MAC-SQL~\cite{wang2023mac} adopts a multi-agent framework where Decomposer, Selector, and Refiner iteratively improve SQL generation. SuperSQL~\cite{li2024dawn} performs automated architecture search across schema linking, prompting, and post-processing with execution-based revision. MCS-SQL~\cite{lee2025mcs} uses multiple prompt templates and a multiple-choice selector to improve validity, while CHESS~\cite{talaei2024chess} and CHASE-SQL~\cite{pourreza2024chase} employ modular, multi-path reasoning pipelines with large models to refine SQL candidates.

\textit{Open-source Fine-tuning Methods.}
T5-3B + PICARD~\cite{scholak2021picard} enforces syntactic validity via constrained decoding. CodeS~\cite{li2024codes} incrementally pretrains open-source models (1B–15B) on curated NL2SQL data to improve structural robustness. SFT Llama2~\cite{li2024can} provides a supervised fine-tuning baseline without reinforcement signals. Alpha-SQL~\cite{li2025alphasql} combines MCTS with LLM reasoning for zero-shot SQL generation. CHESS (open-source)\cite{talaei2024chess} adapts its modular framework to open-source LLMs, while Reasoning-SQL\cite{pourreza2025reasoning} introduces SQL-specific partial rewards for RL. SQL-R1~\cite{ma2025sql} uses coarse reward signals without fine-grained structural feedback, and ExCoT~\cite{zhai2025excot} applies iterative preference optimization guided by execution.

\begin{table}[t]
    \centering
\caption{Comparison of Text-to-SQL methods on BIRD and Spider Dev.}
\label{table:bird_spider_comparison}
\vspace{-1em}
\centering
\resizebox{0.9\columnwidth}{!}{
\begin{tabular}{|l|l|l|}
\hline
\textbf{Methods} & \textbf{BIRD Dev EX\%} & \textbf{Spider Dev EX\%} \\ \hline
\multicolumn{3}{|c|}{\textbf{Proprietary Models ($>$ 70B)}} \\ \hline
DIN-SQL + GPT-4 & 50.72 & 82.8 \\ \hline
DAIL-SQL + GPT-4 & 54.76 & 83.1 \\ \hline
MAC-SQL + GPT-4 & 59.59 & 86.8 \\ \hline
SuperSQL + GPT-4 & 58.50 & 87.0 \\ \hline
MCS-SQL + GPT-4 & 63.36 & 86.8 \\ \hline
CHESS + GPT-4 & 65.00 & -- \\ \hline
CHASE-SQL + Gemini 1.5 & 73.01 & -- \\ \hline
\multicolumn{3}{|c|}{\textbf{Open-source Models ($>$ 5B parameters)}} \\ \hline
SFT Llama2-7B & 45.37 & 77.8 \\ \hline
SFT Llama2-13B & 53.91 & 81.6 \\ \hline
CodeS-7B & 57.17 & 85.4 \\ \hline
CodeS-15B & 58.47 & 84.9 \\ \hline
CHESS (open-source, 33B+70B) & 59.86 & -- \\ \hline
Alpha-SQL 7B & 66.80 & 84.0 \\ \hline
Alpha-SQL 14B & 68.70 & 87.0 \\ \hline
Alpha-SQL 32B & 69.70 & -- \\ \hline
ExCoT 70B & 68.51 & -- \\ \hline
Reasoning-SQL 7B & 64.01 & 78.7 \\ \hline
Reasoning-SQL 14B & 65.31 & 81.4 \\ \hline
SQL-R1 7B & 63.10 & 84.5 \\ \hline
SQL-R1 14B & 67.10 & 86.7 \\ \hline
\multicolumn{3}{|c|}{\textbf{Open-source Models ($\leq$ 5B parameters)}} \\ \hline
Fine-tuned T5-3B + PICARD & 23.34 & 79.3 \\ \hline
CodeS-1B & 50.46 & 77.9 \\ \hline
CodeS-3B & 55.02 & 83.4 \\ \hline
Reasoning-SQL 3B & 58.67 & -- \\ \hline
SQL-R1 3B & 54.6 & 78.1 \\ \hline
\toolname 0.5B (Ours) & 50.85 & 70.2 \\ \hline
\toolname 1.5B (Ours) & 63.17 & 80.0 \\ \hline
\toolname 3B (Ours) & \textbf{67.73} & \textbf{85.0} \\ \hline
\end{tabular}}
\vspace{-1.5em}
\end{table}

\sstitle{Setup} 
We train three policy models—Qwen2.5-Coder-0.5B, 1.5B, and 3B—using GRPO on two NVIDIA A6000 GPUs (48~GB each). All models are trained in bfloat16 precision with Flash Attention~\cite{dao2022flashattention} for efficiency. 
For supervised fine-tuning, the model is trained for two epochs on the reasoning bank in~\autoref{sec:distillation}, with a learning rate of $2 \times 10^{-5}$. 
For GRPO training, the number of rollouts per sample is 32, the learning rate is $8 \times 10^{-6}$, and the global batch size is 32. For the memory reward, we set the number of retrieved reasoning traces 20. With this configuration, the total training time including supervised finetuning and reinforcement learning takes less than 2 days for a 3B model and much faster for 1.5B and 0.5B.

During inference, \toolname follows a lightweight three-stage pipeline comprising schema filtering $\rightarrow$ SQL candidate generation $\rightarrow$ majority voting. This setup is also adopted in many recent systems such as CHESS, CHASE-SQL, and SQL-R1. 
For schema filtering, we employ our prior work GRAST-SQL 0.6B~\footnote{https://huggingface.co/griffith-bigdata/GRAST-SQL-0.6B-BIRD-Reranker}, a schema ranking model that retrieves the top-30 most relevant columns for each query. Unless otherwise specified, all main experiments for accuracy and performance comparison generate 30 SQL candidates per question.

\begin{table}[t]
\centering
\caption{Inference efficiency and deployment cost comparison of Text-to-SQL models on BIRD Dev.}
\label{tab:inference_efficiency}
\vspace{-1em}
\resizebox{0.9\columnwidth}{!}{
\begin{tabular}{|l|c|c|c|l|}
\hline
\textbf{Model} &
\shortstack{\textbf{Latency(s)}} &
\shortstack{\textbf{DType}} &
\shortstack{\textbf{Min. VRAM}\\\textbf{(GB)}} &
\shortstack{\textbf{Compute Setup}\\\textbf{(GPU Memory)}} \\ \hline
\multicolumn{5}{|c|}{\textbf{Closed-Source API-Based Methods}} \\ \hline
DIN-SQL + GPT-4 & 24.09 & -- & -- & OpenAI-based \\
MAC-SQL + GPT-4 & 24.64 & -- & -- & OpenAI-based \\
CHESS$_{(IR, SS, CG)}$ & 118.61 & -- & -- & OpenAI-based \\
CHESS$_{(IR, CG, UT)}$ & 156.50 & -- & -- & OpenAI-based \\ \hline
\multicolumn{5}{|c|}{\textbf{Open-Source Fine-Tuned Models}} \\ \hline
Alpha-SQL 7B & 1650.00 & bf16 & $\sim$22 & 2$\times$46\,GB L40S \\
Alpha-SQL 14B & 1802.00 & bf16 & $\sim$38 & 2$\times$46\,GB L40S \\
Alpha-SQL 32B & 2512.00 & bf16 & $\sim$80 & 4$\times$46\,GB L40S \\
CodeS-1B & 0.69 & fp32 & $\sim$3 & 1$\times$24\,GB A5000 \\
CodeS-3B & 1.06 & fp32 & $\sim$8 & 1$\times$24\,GB A5000 \\
CodeS-7B & 1.87 & fp32 & $\sim$17 & 1$\times$24\,GB A5000 \\
CodeS-15B & 3.52 & fp32 & $\sim$27 & 1$\times$48\,GB A6000 \\ \hline
\multicolumn{5}{|c|}{\textbf{Ours: \toolname}} \\ \hline
\toolname 0.5B & 2.60 & bf16 & $\sim$3 & 1$\times$24\,GB A5000 \\
\toolname 1.5B & 3.25 & bf16 & $\sim$6 & 1$\times$24\,GB A5000 \\
\toolname 3B & 5.57 & bf16 & $\sim$10 & 1$\times$24\,GB A5000 \\ \hline
\end{tabular}
}
\vspace{-2em}
\end{table}

\subsection{End-to-end Comparison}
\label{sec:end2end}

\autoref{table:bird_spider_comparison} presents the execution accuracy (EX\%) of \toolname compared with recent Text-to-SQL systems on the BIRD and Spider development sets. 
Despite its small parameter scale, \toolname consistently achieves accuracy competitive with or superior to much larger models across both benchmarks.
On the BIRD benchmark, \toolname~3B reaches 67.73\% EX, surpassing CodeS-15B (58.47\%), CHESS~(33B+70B, 59.86\%), and Reasoning-SQL~14B (65.31\%), while remaining comparable to Alpha-SQL~14B (68.70\%) and SQL-R1~14B (67.10\%), both of which are over four times larger. 
The 1.5B variant of \toolname achieves 63.17\% EX, outperforming CodeS-7B (57.17\%) and MAC-SQL~(59.59\%) even though these use larger or proprietary backbones. 
On the Spider benchmark, \toolname~3B attains 85.0\% EX, comparable to CodeS-15B (84.9\%), SQL-R1~7B (84.5\%), and Alpha-SQL~7B (84.0\%), and approaching the best-performing 14B-32B models (86--87\%). 
These results demonstrate that our fine-grained execution feedback and dense reward optimization effectively enable small models to close the gap with large-scale systems in both reasoning accuracy and generalization.

Overall, \toolname substantially reduces the dependency on massive model sizes or proprietary APIs. 
With maximum at 3B parameters, it achieves performance on par with or beyond many 14B-70B open-source models, confirming that fine-grained execution and structural feedback provide a cost-efficient path to high Text-to-SQL accuracy.

\subsection{Performance Evaluation}
\label{sec:performance}
\autoref{tab:inference_efficiency} presents the inference latency and deployment cost  of Text-to-SQL models evaluated on the BIRD Dev. 
We compare three system categories: closed-source API-based methods, (ii) open-source fine-tuned models, and (iii) our proposed \toolname family. 
All open-source systems are deployed using vLLM \cite{kwon2023efficient}, 
where the minimum VRAM required for serving LLMs can be estimated as the sum of model parameters, KV cache, and minor ephemeral memory:
\[
\text{VRAM}_{\text{serve}} \;\approx\; 
\text{VRAM}_{\text{model}} \;+\; 
\text{VRAM}_{\text{KV-cache}} \;+\; 
\text{VRAM}_{\text{others}} .
\]
with KV-cache scaling linearly with context length (e.g., 40K tokens in Alpha-SQL greatly increase memory use).

While large models such as Alpha-SQL 14B and Alpha-SQL 32B achieve slightly higher execution accuracy (\autoref{table:bird_spider_comparison}), they incur prohibitive inference latency (1800-2500s/sample) and memory costs (up to $\sim$80GB VRAM). In contrast, \toolname3B attains comparable accuracy with much lower overhead (5.57s/sample, $\sim$10GB VRAM). Proprietary pipelines such as CHESS remain competitive in accuracy but suffer from multi-stage inference latency exceeding 100s/sample due to sequential API calls~\cite{talaei2024chess,chung2025long}, making them unsuitable for latency-sensitive or on-premise settings.

Importantly, the compact memory footprint of \toolname enables deployment on cost-effective GPUs. 
The 3B-parameter variant can operate comfortably on widely available and affordable consumer GPUs, such as the NVIDIA RTX 5070 (12 GB), RTX 4090 or A5000 (24 GB), without requiring specialized server-grade hardware.
This substantially lowers the barrier to real-world adoption compared to large-scale models requiring multi-GPU racks or proprietary API.

Overall, these results show that \toolname achieves a strong balance between accuracy, latency, and cost, enabling practical Text-to-SQL deployment on commodity hardware.
Note that this analysis includes only models with public implementations; methods such as CHASE-SQL~\cite{pourreza2024chase} and SQL-R1~\cite{ma2025sql} lack open-source code, preventing fair measurement of inference time under comparable settings.

\begin{figure}[t]
	\centering
	\includegraphics[width=0.85\linewidth]{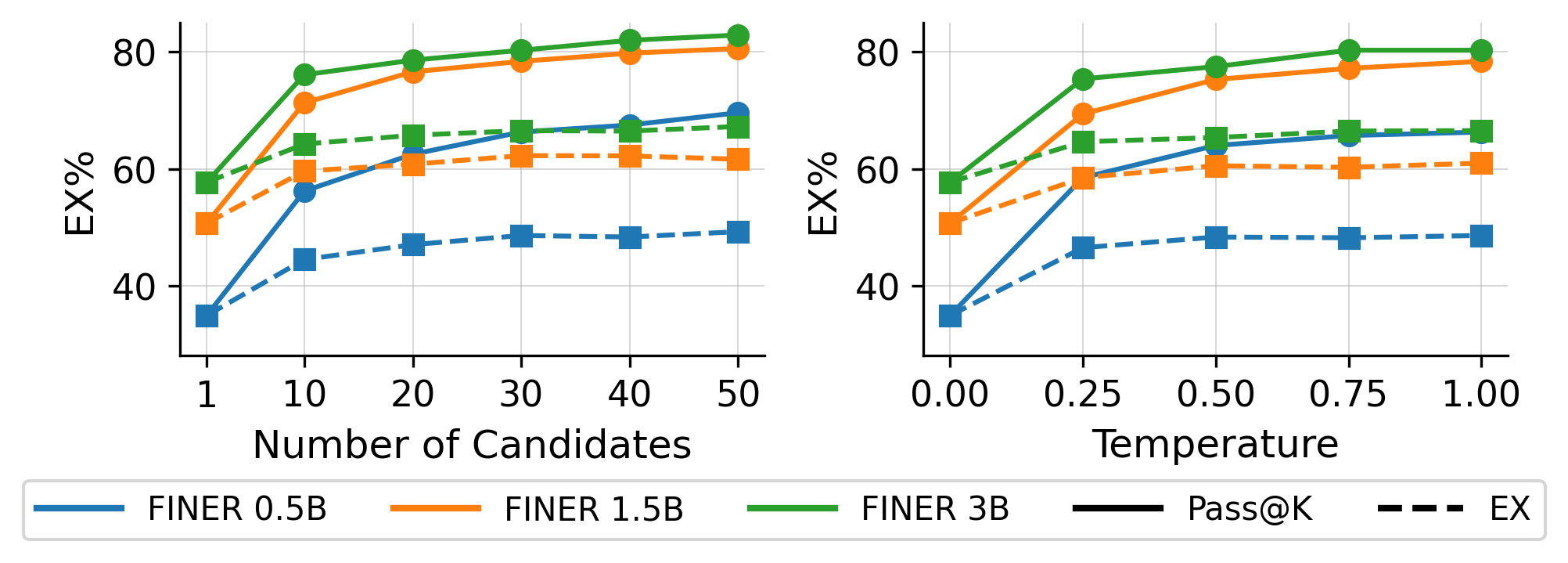}
    \vspace{-1.em}
	\caption{\toolname Pass@K and EX\% on BIRD Dev under different numbers of candidates (left) and sampling temperatures (right).}
	\label{fig:selection_effect}
    \vspace{-1em}
\end{figure}

\begin{figure}[t]
    \centering
    \setlength{\tabcolsep}{0pt}
    \begin{tabular}{ccc}
        \includegraphics[width=0.34\linewidth]{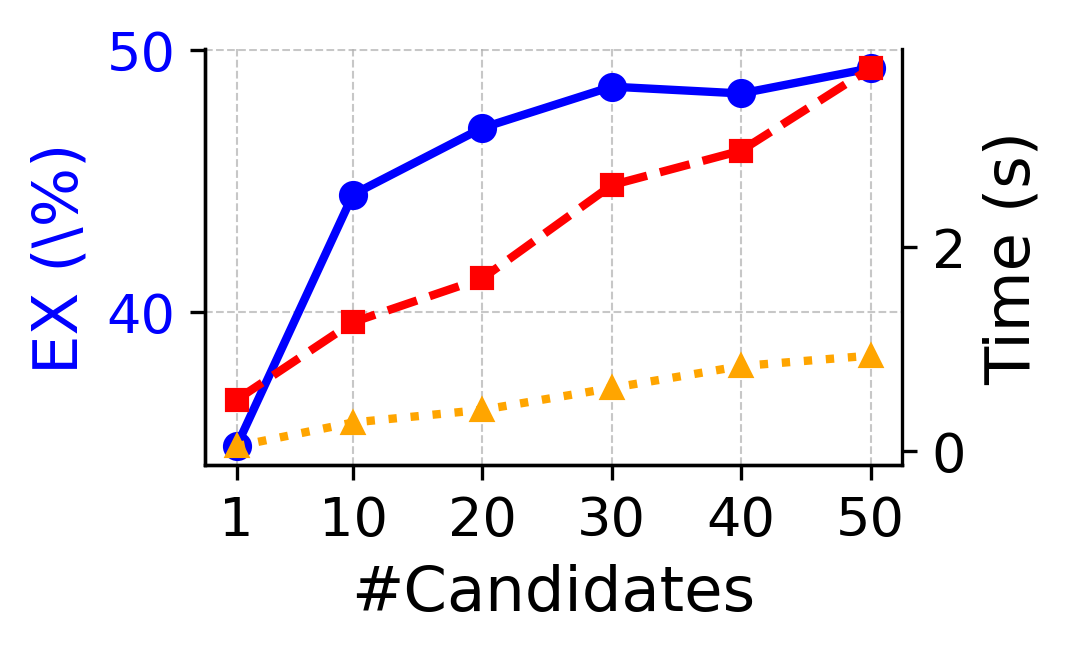} &
        \includegraphics[width=0.34\linewidth]{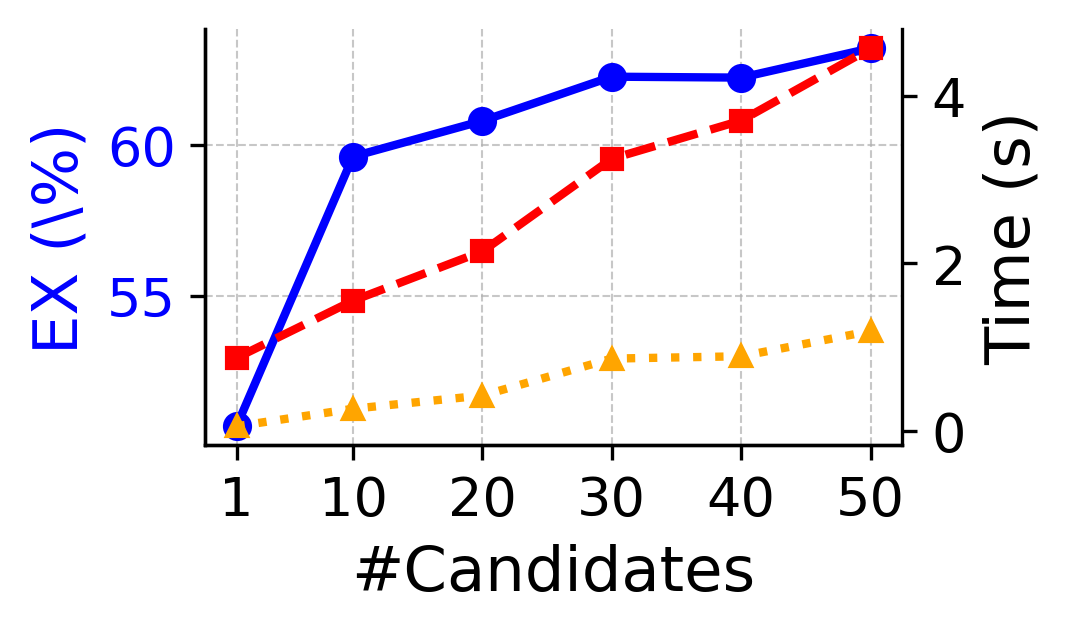} &
        \includegraphics[width=0.34\linewidth]{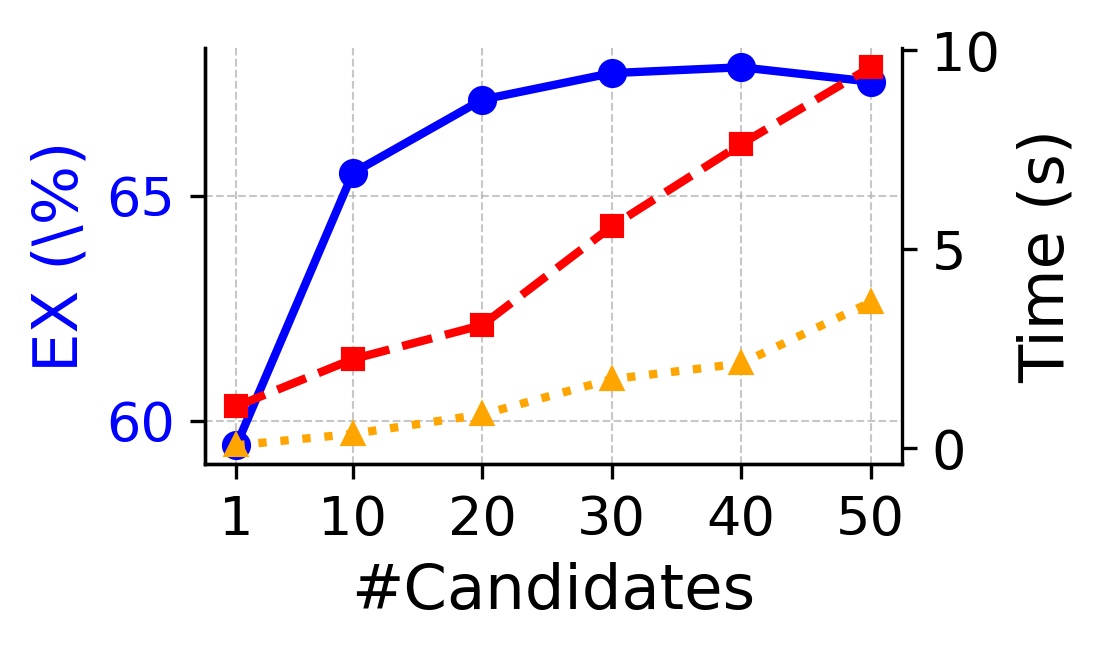} \\
        \small (a) 0.5B &
        \small (b) 1.5B &
        \small (c) 3B \\
    \end{tabular}
    \caption{Accuracy--latency trade-off of candidate selection for \toolname\ (left-to-right: 0.5B, 1.5B, 3B). The blue curve denotes EX\%, the red curve denotes end-to-end average latency per query, and the yellow curve denotes execution-only latency.}
    \label{fig:accuracy_latency_tradeoff}
    \vspace{-2em}
\end{figure}

\subsection{Hyperparameter Sensitivity}
\label{sec:sensitivity}

\sstitle{Number of candidates}
We analyze the impact of the number of sampled SQL candidates $n$ on inference performance, as shown in \autoref{fig:selection_effect} (left). Increasing $n$ from 1 to 50 consistently improves both Pass@K and EX\% across all FINER models, but the gain saturates beyond 30 candidates. The smallest model (0.5B) benefits the most, rising from 35\% to nearly 49\% EX, indicating that stochastic sampling and majority voting effectively compensate for limited capacity. Larger models (1.5B and 3B) exhibit higher initial accuracy and more stable convergence, reaching around 82\% Pass@K and 68\% EX at $n{=}50$. These findings demonstrate that moderate candidate sampling provides an efficient trade-off between performance gains and computational cost during inference.

We show the accuracy versus latency trade-off in \autoref{fig:accuracy_latency_tradeoff}. As the number of candidates increases, EX\% (blue) improves steadily, but it comes with the increasing in end-to-end latency (red) and SQL execution latency. Across model scales, $n$ in the range of 20--30 provides the best operating point, balance between accuracy and latency. This analysis guides the choice of $n$ in interactive settings, noting that absolute execution time depends on database and SQL complexity.

\sstitle{Temperature for candidate generation}
We further analyze the impact of sampling temperature on SQL diversity and execution accuracy. As illustrated in \autoref{fig:selection_effect} (right), raising the temperature from 0 to 1.0 consistently enhances both Pass@K and EX\% across all model sizes. Higher temperatures encourage exploration of diverse reasoning and SQL structures, which leads to more correct candidates under majority voting. The smallest model (0.5B) shows the strongest improvement, with EX increasing from 35\% at greedy decoding to nearly 49\% at temperature~1.0. Larger models (1.5B and 3B) also benefit moderately, stabilizing around 60-68\% EX. Overall, the best configuration is achieved at temperature~1.0 with 50 sampled candidates, showing that moderate stochastic sampling effectively balances exploration and accuracy.

\begin{figure}[t]
    \centering
    \includegraphics[width=0.78\linewidth]{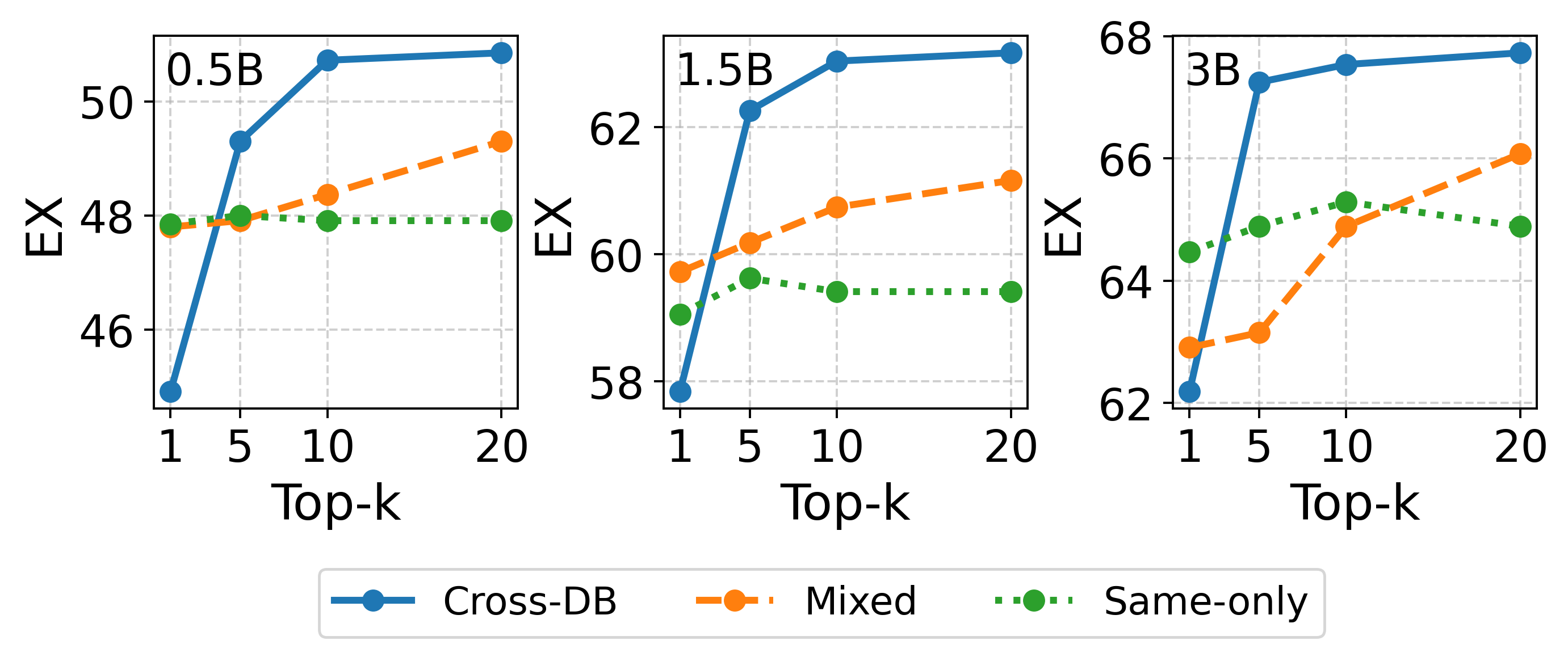}
    \vspace{-1em}
    \caption{Sensitivity of memory retrieval top-$k$ on BIRD Dev EX\% with three retrieval scopes: Cross-DB (retrieving only from other databases), Same-only (retrieval only from current database), and Mixed (query from all databases).}
    \label{fig:mem_topk_sensitivity}
    \vspace{-1.8em}
\end{figure}

\sstitle{Memory configuration}
\autoref{fig:mem_topk_sensitivity} analyzes the effect of memory configuration by varying \#retrieved reasoning traces ($k$) and the retrieval scope. Across all model sizes, increasing $k$ yields modest gains that saturate around $k\in\{10,20\}$, while Cross-DB retrieval consistently achieves the highest EX\% in the high-$k$ regime. By excluding the querying database, Cross-DB retrieval enables access to more diverse reasoning traces from different databases. In contrast, in the Mixed setting, retrieval is often dominated by traces from the same database due to strong surface-level entity overlap, causing behavior and performance close to the Same-only setting. These results indicate that cross-database diversity is crucial for effective memory-based supervision, taking its centroid embedding allows transfer the reasoning patterns to SLMs.

\begin{figure}[!h]
    \centering
    \vspace{-1em}
    \includegraphics[width=0.65\linewidth]{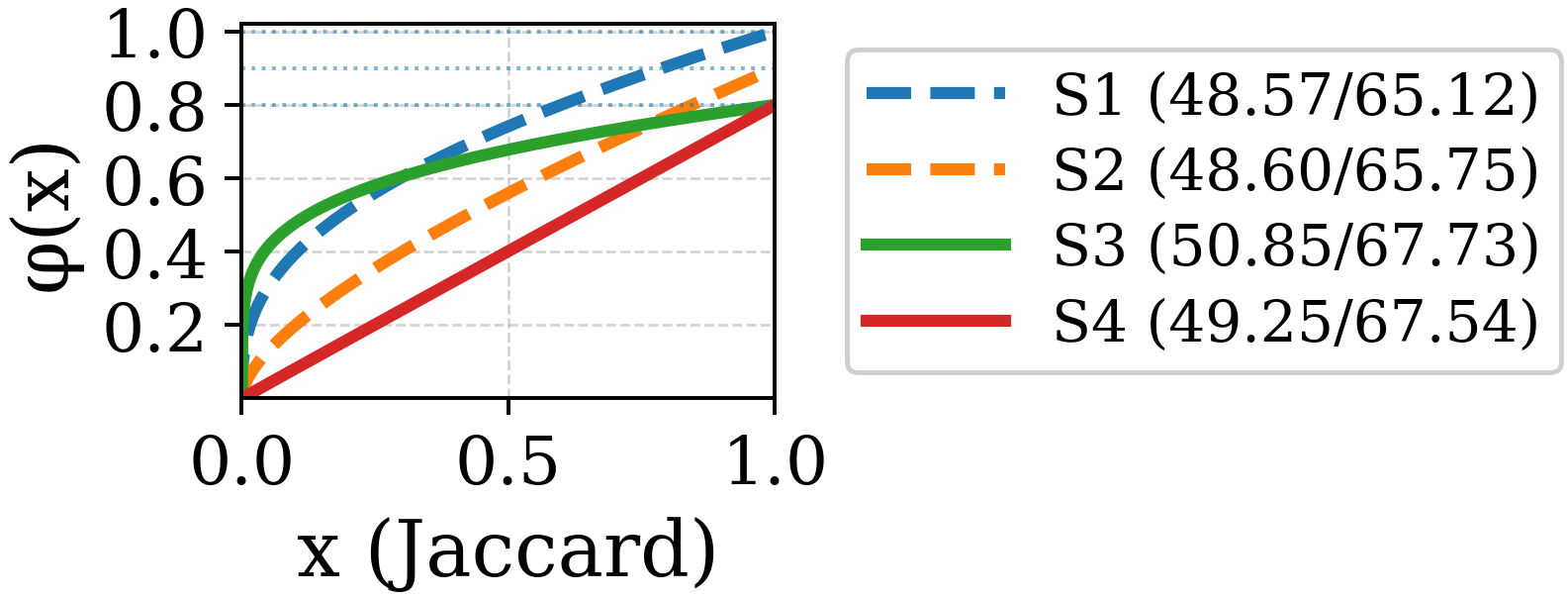}
    \vspace{-1em}
    \caption{Sensitivity of the shaping function in Atomic Reward under four settings. The legend reports EX on BIRD Dev for 0.5B/3B as EX$_{0.5B}$/EX$_{3B}$.}
    \label{fig:atomic_shaping_sensitivity}
    \vspace{-0.8em}
\end{figure}

\sstitle{Atomic Shaping Function Sensitivity}
We analyze the sensitivity of the atomic-op shaping function
$\varphi(x)=\lambda\,x+(1-\lambda)\beta x^\gamma$ as introduced in \autoref{sec:atomic_reward}.
Our design objective is to allocate relatively higher reward to low-overlap candidates (larger $\varphi(x)$ when $x\!\approx\!0$) while preventing reward saturation for near-perfect matches by controlling the endpoint $\varphi(1)$ (smaller $\varphi(x)$ when $x\!\approx\!1$).
To keep the sweep compact, we select four representative settings by inspecting the geometry of $\varphi(x)$ and covering both endpoint caps and concavity:
S1 $(\lambda{=}0.15,\beta{=}1.0,\gamma{=}0.35)$,
S2 $(\lambda{=}0.30,\beta{=}0.85,\gamma{=}0.55)$,
S3 $(\lambda{=}0.05,\beta{=}0.79,\gamma{=}0.20)$,
and S4 $(\lambda{=}0.60,\beta{=}0.50,\gamma{=}0.98)$ which is near linear.
\autoref{fig:atomic_shaping_sensitivity} visualizes the resulting shaping curves, and the legend reports the corresponding execution accuracy on BIRD Dev for 0.5B and 3B (EX$_{0.5B}$/EX$_{3B}$). 
S3 and S4 achieve the best EX across both 0.5B and 3B models, and both share a capped endpoint $\varphi(1)\approx0.8$, which suppresses the reward for near-perfect atomic overlap when execution is not fully correct.
Notably, S3 assigns much higher reward near Jaccard score near 0 and yields the strongest gains on the 0.5B model, reinforcing our design choice to boost low-overlap candidates while suppressing Jaccard score near 1.

\subsection{Performance Gains through GRPO Training Steps}
\label{sec:alignment_vs_sft}

\begin{figure}[!t]
	\centering
	\includegraphics[width=0.7\linewidth]{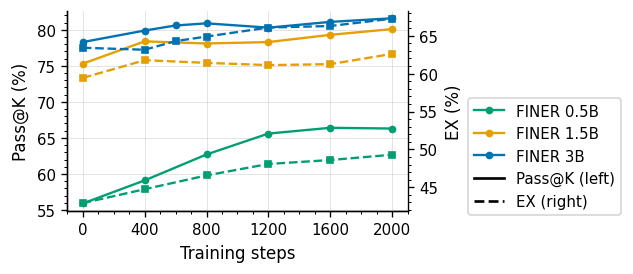}
	\vspace{-1.0em}
	\caption{Performance improvement of Pass@K and EX on BIRD~Dev across GRPO training steps. 
	Step 0 represents the supervised fine-tuning baseline.}
	\label{fig:rlef_iterations}
	\vspace{-1.0em}
\end{figure}

We analyze the impact of GRPO optimization over successive training steps, as illustrated in \autoref{fig:rlef_iterations}. The curves report Pass@K and EX\% on BIRD~Dev, where step~0 corresponds to the supervised fine-tuned baseline and subsequent points reflect progressive policy updates guided by dense execution feedback. All models exhibit consistent improvement, confirming that GRPO effectively refines reasoning and SQL generation through iterative interaction with the environment. The 0.5B model gains the most, improving from 55.9\% to 66.3\% Pass@K and from 42.9\% to 49.3\% EX after 2000 steps. Larger models (1.5B and 3B) also show steady improvements, with the 1.5B model rising from 75.3\%/59.4\% to 80.1\%/63.1\% and the 3B model from 78.3\%/63.4\% to 81.6\%/67.7\% (Pass@K/EX). These results highlight that GRPO enables stable and monotonic performance gains across different model scales, with smaller models benefiting the most from fine-grained execution feedback.

\begin{figure}[t]
    \centering
    \includegraphics[width=0.9\linewidth]{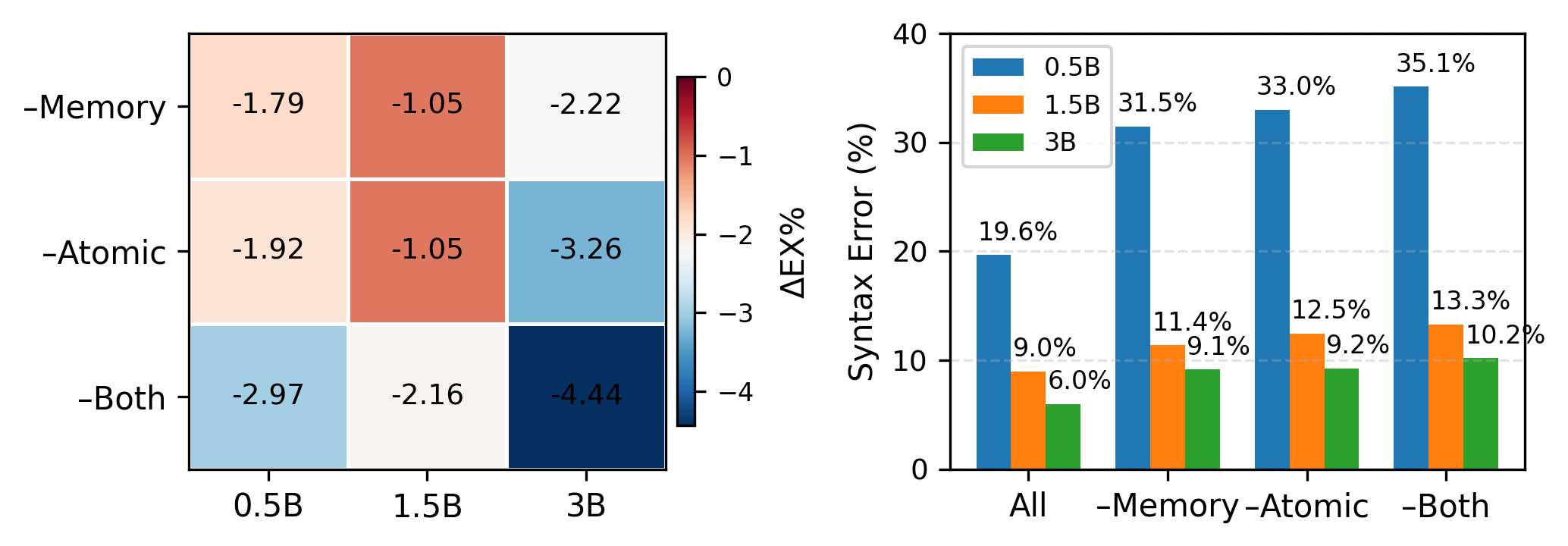}
    \vspace{-1.0em}
    \caption{Ablation of reward components. Left: $\Delta$EX\% when removing each reward. Right: syntax error rate when removing reward components.}
    \label{fig:ablation}
    \vspace{-1.5em}
\end{figure}

\subsection{Ablation Study}
\label{sec:ablation}

\paragraph{Impact of Reward Components}
We evaluate the contribution of the memory and atomic rewards by removing them individually or jointly during GRPO training.  
As shown in \autoref{fig:ablation}, excluding either reward leads to a clear drop in EX\% and a rise in syntax error rates across all model sizes.  
For the 0.5B model, removing the memory reward decreases EX by 1.79\% and raises syntax errors from 19.6\% to 31.5\%, while removing the atomic reward causes a similar 1.92\% EX drop.  
For the 3B model, removing the memory or atomic reward reduces EX by 2.22\% and 3.26\%, respectively, with syntax errors increasing from 6.0\% to about 10\%.  
When both rewards are removed, degradation becomes most severe (-2.97\%, -2.16\%, and -4.44\% EX for 0.5B, 1.5B, and 3B), confirming their complementary effects.  

Removing the memory reward causes higher syntax error rates, as the model explores unstable reasoning paths that lead to invalid or unexecutable SQLs.  
Eliminating both rewards further lowers EX due to the sparse reward problem—when failures yield zero feedback, the model stops improving over time.  
Overall, the memory reward stabilizes reasoning and reduces invalid SQLs, while the atomic reward enforces operation-level structural correctness.  
Together, they provide fine-grained signals that guide learning and enable the model to improve even when the generated SQL is incorrect.

\begin{table}[t]
    \centering
    \caption{Reasoning Stability Analysis: No Memory vs. With Memory.}
    \label{tab:ablation_stability}
    \vspace{-.5em}
    \resizebox{1\columnwidth}{!}{
    \begin{tabular}{|l|cc|cc|cc|}
    \hline
    \multirow{2}{*}{\textbf{Metrics}} & \multicolumn{2}{c|}{\textbf{\toolname 0.5B}} & \multicolumn{2}{c|}{\textbf{\toolname 1.5B}} & \multicolumn{2}{c|}{\textbf{\toolname 3B}} \\ \cline{2-7}
     & \textbf{w/o Mem} & \textbf{w/ Mem} & \textbf{w/o Mem} & \textbf{w/ Mem} & \textbf{w/o Mem} & \textbf{w/ Mem} \\ \hline
    EX (\%) & 46.28 & \textbf{50.85} & 61.21 & \textbf{63.17} & 65.38 & \textbf{67.73} \\ \hline
    Self-BLEU & 0.674 & \textbf{0.724} & 0.607 & \textbf{0.631} & 0.584 & \textbf{0.615} \\ \hline
    Mean Exec. Groups & 8.07 & \textbf{5.59} & 7.16 & \textbf{6.25} & 6.21 & \textbf{5.27} \\ \hline
    \end{tabular}}
    \vspace{-1.0em}
\end{table}

\begin{figure}[t]
  \centering
  \includegraphics[width=0.9\columnwidth]{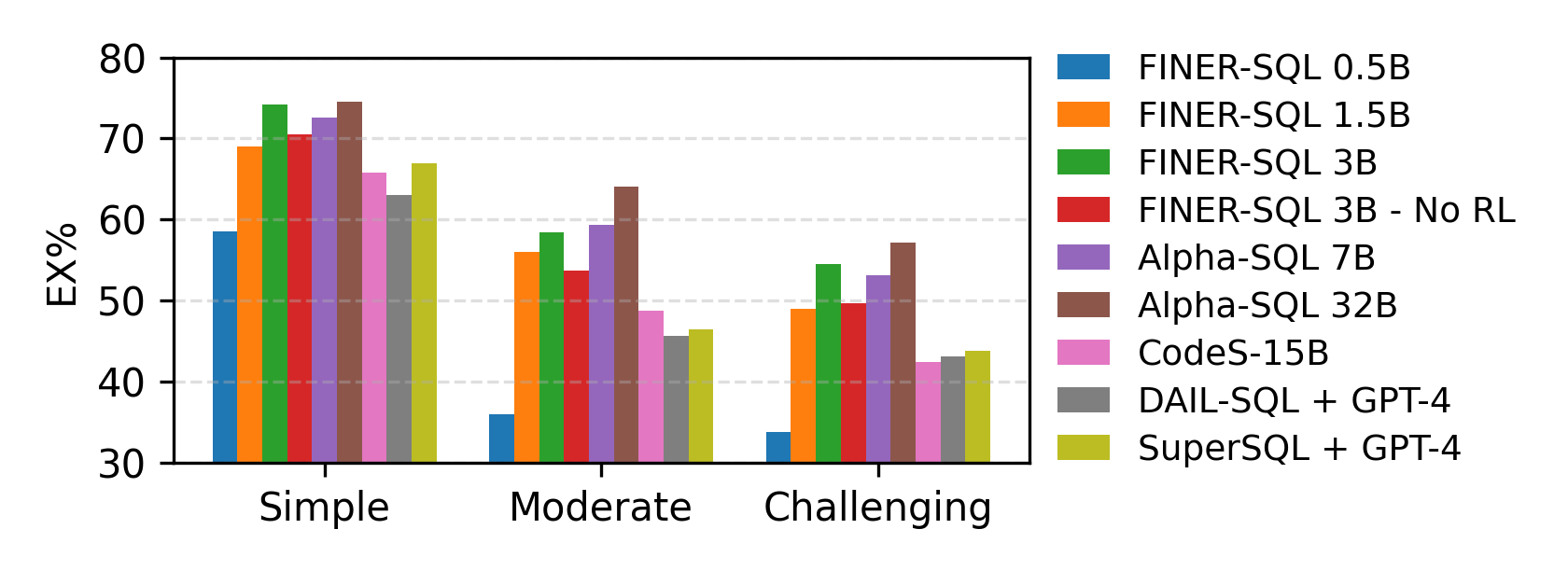}
  \vspace{-1.5em}
  \caption{EX\% by SQL complexity on BIRD Dev, showing \toolname (1.5B-3B) stay competitive on challenging queries.}
  \label{fig:complexity_barchart}
  \vspace{-1.5em}
\end{figure}

\paragraph{Analysis of Reasoning Stability}
Given each prompt, we sample $K{=}30$ candidate reasoning traces and their corresponding SQL queries.
We quantify the diversity of the reasoning texts using Self-BLEU \cite{zhu2018texygen}, where each trace is treated as the hypothesis and the remaining $K{-}1$ traces serve as references; thus, a higher Self-BLEU indicates lower diversity.
\autoref{tab:ablation_stability} shows that enabling memory consistently increases Self-BLEU, meaning the model produces less diverse reasoning across candidates.
This reduced reasoning variance propagates to the generated SQL, yielding fewer distinct execution results (lower mean execution groups), which is beneficial for majority voting: when candidates concentrate into fewer execution clusters, votes are less fragmented and the correct execution is more likely to win.
For instance, at 0.5B, memory raises Self-BLEU from 0.674 to 0.724 while reducing mean execution groups from 8.07 to 5.59, improving execution accuracy from 46.28\% to 50.85\%.

\subsection{Evaluation on SQL Complexity}
\label{sec:complexity}

To examine how model scale and our reward design influence reasoning robustness, we analyze performance across three SQL complexity levels -- \emph{simple}, \emph{moderate}, and \emph{challenging} on the BIRD Dev.
\autoref{fig:complexity_barchart} presents results for all competing systems.
Despite relying on SLMs, \toolname maintains strong and consistent performance as query complexity increases.  
The 3B variant reaches 67.7\% EX, closely approaching Alpha-SQL 32B and surpassing Alpha-SQL 7B.
Meanwhile, the 1.5B model achieves 56.0\% EX on moderate and 49.0\% EX on challenging queries—outperforming DAIL-SQL and SuperSQL with GPT-4.
We further observe that RL is a key driver of this robustness.
Comparing \toolname 3B against its \emph{No RL} counterpart, RL consistently improves execution accuracy across all complexity levels:
74.16 vs.\ 70.49 on \emph{simple}, 58.41 vs.\ 53.66 on \emph{moderate}, and 54.48 vs.\ 49.66 on \emph{challenging}.
These results also suggest that methodology can outweigh raw model scale: even when using SLMs, \toolname can handle complex reasoning and compositional SQL generation effectively, and can rival or exceed much larger or GPT-4-assisted baselines.

\begin{figure}[t]
    \centering
    \includegraphics[width=0.7\linewidth]{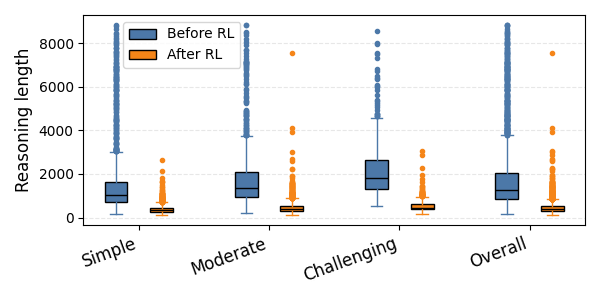}
    \vspace{-1.5em}
    \caption{
    The reasoning-length distributions before and after RL across query difficulties. 
    Our RL framework leads to more concise reasoning traces.
    }
    \label{fig:reasoning_lengths}
    \vspace{-1.5em}
\end{figure}

\begin{tcolorbox}[colback=gray!20, colframe=gray!90, boxrule=0.3pt, arc=1pt, left=3pt, right=3pt, top=2pt, bottom=2pt, breakable]
\tiny
\textbf{Example of reasoning before RL: 7.8K characters \& incorrect}\\
First, the user wants: "list the lowest three eligible free rates for students aged 5-17 in continuation schools."\\
From the schema, the eligible free rate is in frpm.`Percent (\%) Eligible Free (Ages 5-17)`. The condition "continuation schools" refers to School Type = 'Continuation High Schools'. External knowledge says: eligible free rate = Free Meal Count (Ages 5-17) / Enrollment (Ages 5-17). I should check which column represents that ratio... it may already exist as a computed field.\\
...\\
Maybe I need to join with schools to confirm the school type or use frpm.School Type.\\
To be accurate, I should select from frpm where School Type = 'Continuation High Schools', order by the eligible free rate ascending, and limit to 3 results.
\end{tcolorbox}

\begin{tcolorbox}[colback=gray!20, colframe=gray!90, boxrule=0.3pt, arc=1pt, left=3pt, right=3pt, top=2pt, bottom=2pt, breakable]
\tiny
\textbf{Example of reasoning after RL: concise \& correct SQL}\\
We need the three lowest eligible free rates for students under 18, defined by frpm.`Free Meal Count (Ages 5-17)` / frpm.`Enrollment (Ages 5-17)` for schools where frpm.`School Type` is "Continuation High Schools". Return rates. Note we must compute the rate even if there are nulls: SQL will ignore them. Order by rate ascending, limit 3. Use explicit CTE or subquery for clarity. Return only the rate.
\end{tcolorbox}

\subsection{Reasoning Evolves under RL}
\label{sec:reasoning_length}

To assess how \toolname shapes reasoning behavior, we compare the 3B model before and after RL training.
\autoref{fig:reasoning_lengths} shows reasoning-length distributions across query difficulties.
Before RL, the model produces long, inconsistent chains
(\textit{Simple}: 2.38K, \textit{Moderate}: 2.93K, \textit{Challenging}: 3.49K on average, many exceeding 8K),
reflecting redundant exploration and schema confusion that often fails to converge.
After RL, reasoning is substantially shorter and more stable
(\textit{Simple}: 0.57K, \textit{Moderate}: 0.74K, \textit{Challenging}: 0.85K),
indicating a shift toward concise, goal-oriented reasoning that yields correct SQL.
A representative case from the \textit{california\_schools} database highlights this shift.
\textit{Before RL} (7.8K characters), the model redundantly restates schema elements, explores irrelevant joins, and fails to produce a correct query.
\textit{After RL} (0.7K characters), it quickly identifies relevant tables and fields, applies the correct expression and filter, and completes the SQL concisely.

\section{Conclusion}
\label{sec:con}

We presented \toolname, a reinforcement learning framework that boosts small language models for Text-to-SQL through fine-grained execution feedback and cost-efficient rewards.
With dense and interpretable signals (memory and atomic rewards), \toolname turns sparse supervision into continuous guidance, enabling stable GRPO optimization even when SQL fails.
Experiments on Spider and BIRD show that 1.5B-3B models trained with our method achieve competitive execution accuracy with 14B-70B models at a fraction of the cost, offering a practical path for efficient and privacy-preserving Text-to-SQL deployment.



\begin{thebibliography}{10}
\providecommand{\url}[1]{#1}
\csname url@samestyle\endcsname
\providecommand{\newblock}{\relax}
\providecommand{\bibinfo}[2]{#2}
\providecommand{\BIBentrySTDinterwordspacing}{\spaceskip=0pt\relax}
\providecommand{\BIBentryALTinterwordstretchfactor}{4}
\providecommand{\BIBentryALTinterwordspacing}{\spaceskip=\fontdimen2\font plus
\BIBentryALTinterwordstretchfactor\fontdimen3\font minus
  \fontdimen4\font\relax}
\providecommand{\BIBforeignlanguage}[2]{{%
\expandafter\ifx\csname l@#1\endcsname\relax
\typeout{** WARNING: IEEEtran.bst: No hyphenation pattern has been}%
\typeout{** loaded for the language `#1'. Using the pattern for}%
\typeout{** the default language instead.}%
\else
\language=\csname l@#1\endcsname
\fi
#2}}
\providecommand{\BIBdecl}{\relax}
\BIBdecl

\bibitem{wang2020rat}
B.~Wang, R.~Shin, X.~Liu, O.~Polozov, and M.~Richardson, ``Rat-sql:
  Relation-aware schema encoding and linking for text-to-sql parsers,'' in
  \emph{ACL}, 2020, pp. 7567--7578.

\bibitem{chang2023drspider}
S.~Chang, J.~Wang, M.~Dong, L.~Pan, H.~Zhu \emph{et~al.}, ``Dr.spider: A
  diagnostic evaluation benchmark towards text-to-{SQL} robustness,'' in
  \emph{ICLR}, 2023.

\bibitem{gan-etal-2021-exploring}
Y.~Gan, X.~Chen, and M.~Purver, ``Exploring underexplored limitations of
  cross-domain text-to-{SQL} generalization,'' in \emph{EMNLP}, 2021, pp.
  8926--8931.

\bibitem{hung2019handling}
N.~Q.~V. Hung, M.~Weidlich, N.~T. Tam, Z.~Mikl{\'o}s, K.~Aberer, A.~Gal, and
  B.~Stantic, ``Handling probabilistic integrity constraints in pay-as-you-go
  reconciliation of data models,'' \emph{Information Systems}, vol.~83, pp.
  166--180, 2019.

\bibitem{nguyen2026handling}
M.~H. Nguyen, T.~T. Nguyen, J.~Jo, D.~A. Nguyen, H.~Yin, and Q.~V.~H. Nguyen,
  ``Handling data sparsity and model poisoning attacks in federated sequential
  recommender systems,'' \emph{Knowledge-Based Systems}, p. 115545, 2026.

\bibitem{nguyen2024handling}
T.~T. Nguyen, T.~T. Nguyen, M.~Weidlich, J.~Jo, Q.~V.~H. Nguyen, H.~Yin, and
  A.~W.-C. Liew, ``Handling low homophily in recommender systems with
  partitioned graph transformer,'' \emph{IEEE Transactions on Knowledge and
  Data Engineering}, 2024.

\bibitem{yu-etal-2018-spider}
T.~Yu, R.~Zhang, K.~Yang, M.~Yasunaga, D.~Wang, Z.~Li, J.~Ma, I.~Li, Q.~Yao,
  S.~Roman, Z.~Zhang, and D.~Radev, ``{S}pider: A large-scale human-labeled
  dataset for complex and cross-domain semantic parsing and text-to-{SQL}
  task,'' in \emph{EMNLP}, 2018, pp. 3911--3921.

\bibitem{li2024can}
J.~Li, B.~Hui, G.~Qu, J.~Yang, B.~Li, B.~Li, B.~Wang, B.~Qin, R.~Geng, N.~Huo
  \emph{et~al.}, ``Can llm already serve as a database interface? a big bench
  for large-scale database grounded text-to-sqls,'' \emph{NeurIPS}, vol.~36,
  2024.

\bibitem{li2024codes}
H.~Li, J.~Zhang, H.~Liu, J.~Fan, X.~Zhang, J.~Zhu, R.~Wei, H.~Pan, C.~Li, and
  H.~Chen, ``Codes: Towards building open-source language models for
  text-to-sql,'' \emph{SIGMOD}, vol.~2, no.~3, pp. 1--28, 2024.

\bibitem{talaei2024chess}
S.~Talaei, M.~Pourreza, Y.-C. Chang, A.~Mirhoseini, and A.~Saberi, ``Chess:
  Contextual harnessing for efficient sql synthesis,'' \emph{arXiv preprint
  arXiv:2405.16755}, 2024.

\bibitem{pourreza2024chase}
M.~Pourreza, H.~Li, R.~Sun, Y.~Chung, S.~Talaei, G.~T. Kakkar, Y.~Gan,
  A.~Saberi, F.~Ozcan, and S.~O. Arik, ``Chase-sql: Multi-path reasoning and
  preference optimized candidate selection in text-to-sql,'' \emph{arXiv
  preprint arXiv:2410.01943}, 2025.

\bibitem{pham2026modeval}
K.~T. Pham, T.~T. Nguyen, V.~Huynh, H.~Yin, and Q.~V.~H. Nguyen, ``An efficient
  and effective evaluator for text2sql models on unseen and unlabeled data,''
  in \emph{2026 IEEE 42nd International Conference on Data Engineering
  (ICDE)}.\hskip 1em plus 0.5em minus 0.4em\relax IEEE, 2026.

\bibitem{ren2022prototype}
Z.~Ren, T.~T. Nguyen, and W.~Nejdl, ``{Prototype learning for interpretable
  respiratory sound analysis},'' in \emph{Proc.\ ICASSP}, 2022, pp. 9087--9091.

\bibitem{pham2024dual}
M.~T. Pham, T.~T. Huynh, T.~T. Nguyen, T.~T. Nguyen, T.~T. Nguyen, J.~Jo,
  H.~Yin, and Q.~V. Hung~Nguyen, ``A dual benchmarking study of facial forgery
  and facial forensics,'' \emph{CAAI Transactions on Intelligence Technology},
  vol.~9, no.~6, pp. 1377--1397, 2024.

\bibitem{abedini2025masksql}
S.~Abedini, S.~Mohapatra, D.~Emerson, M.~Shafieinejad, J.~C. Cresswell, and
  X.~He, ``Masksql: Safeguarding privacy for llm-based text-to-sql via
  abstraction,'' \emph{arXiv preprint arXiv:2509.23459}, 2025.

\bibitem{hui2025privacypad}
Z.~Hui, Y.~R. Dong, S.~Sivapiromrat, E.~Shareghi, and N.~Collier, ``Privacypad:
  A reinforcement learning framework for dynamic privacy-aware delegation,''
  \emph{arXiv preprint arXiv:2510.16054}, 2025.

\bibitem{nguyen2025privacy}
T.~T. Nguyen, T.~T. Huynh, Z.~Ren, T.~T. Nguyen, P.~L. Nguyen, H.~Yin, and
  Q.~V.~H. Nguyen, ``Privacy-preserving explainable ai: a survey,''
  \emph{Science China Information Sciences}, vol.~68, no.~1, p. 111101, 2025.

\bibitem{nguyen2023poisoning}
T.~Nguyen~Thanh, N.~D.~K. Quach, T.~T. Nguyen, T.~T. Huynh, V.~H. Vu, P.~L.
  Nguyen, J.~Jo, and Q.~V.~H. Nguyen, ``Poisoning gnn-based recommender systems
  with generative surrogate-based attacks,'' \emph{ACM Transactions on
  Information Systems}, vol.~41, no.~3, pp. 1--24, 2023.

\bibitem{nguyen2023detecting}
T.~T. Nguyen, T.~T. Huynh, H.~Yin, M.~Weidlich, T.~T. Nguyen, T.~S. Mai, and
  Q.~V.~H. Nguyen, ``Detecting rumours with latency guarantees using massive
  streaming data,'' \emph{The VLDB Journal}, vol.~32, no.~2, pp. 369--387,
  2023.

\bibitem{chung2025long}
Y.~Chung, G.~T. Kakkar, Y.~Gan, B.~Milne, and F.~Ozcan, ``Is long context all
  you need? leveraging llm's extended context for {NL2SQL},'' \emph{PVLDB},
  vol.~18, no.~8, pp. 2735--2747, 2025.

\bibitem{touvron2023llama}
H.~Touvron, T.~Lavril, G.~Izacard, X.~Martinet, M.-A. Lachaux, T.~Lacroix,
  B.~Rozi{\`e}re, N.~Goyal, E.~Hambro, F.~Azhar \emph{et~al.}, ``Llama: Open
  and efficient foundation language models,'' \emph{arXiv preprint
  arXiv:2302.13971}, 2023.

\bibitem{jiang2023mistral}
A.~Q. Jiang, A.~Sablayrolles, A.~Mensch \emph{et~al.}, ``Mistral 7b,''
  \emph{arXiv preprint arXiv:2310.06825}, 2023.

\bibitem{huynh2023efficient}
T.~T. Huynh, M.~H. Nguyen, T.~T. Nguyen, P.~L. Nguyen, M.~Weidlich, Q.~V.~H.
  Nguyen, and K.~Aberer, ``Efficient integration of multi-order dynamics and
  internal dynamics in stock movement prediction,'' in \emph{Proceedings of the
  Sixteenth ACM International Conference on Web Search and Data Mining}, 2023,
  pp. 850--858.

\bibitem{duong2022efficient}
C.~T. Duong, T.~T. Nguyen, H.~Yin, M.~Weidlich, T.~S. Mai, K.~Aberer, and
  Q.~V.~H. Nguyen, ``Efficient and effective multi-modal queries through
  heterogeneous network embedding,'' \emph{IEEE Transactions on Knowledge and
  Data Engineering}, vol.~34, no.~11, pp. 5307--5320, 2022.

\bibitem{nguyen2024portable}
T.~T. Nguyen, Z.~Ren, T.~T. Nguyen, J.~Jo, Q.~V.~H. Nguyen, and H.~Yin,
  ``Portable graph-based rumour detection against multi-modal heterophily,''
  \emph{Knowledge-Based Systems}, vol. 284, p. 111310, 2024.

\bibitem{nguyen2024manipulating}
T.~T. Nguyen, N.~Quoc Viet~Hung, T.~T. Nguyen, T.~T. Huynh, T.~T. Nguyen,
  M.~Weidlich, and H.~Yin, ``Manipulating recommender systems: A survey of
  poisoning attacks and countermeasures,'' \emph{ACM Computing Surveys},
  vol.~57, no.~1, pp. 1--39, 2024.

\bibitem{nguyen2015smart}
Q.~V.~H. Nguyen, T.~T. Nguyen, V.~T. Chau, T.~K. Wijaya, Z.~Mikl{\'o}s,
  K.~Aberer, A.~Gal, and M.~Weidlich, ``Smart: A tool for analyzing and
  reconciling schema matching networks,'' in \emph{ICDE}, 2015, pp. 1488--1491.

\bibitem{nguyen2015tag}
Q.~V.~H. Nguyen, S.~T. Do, T.~T. Nguyen, and K.~Aberer, ``Tag-based paper
  retrieval: minimizing user effort with diversity awareness,'' in
  \emph{International Conference on Database Systems for Advanced
  Applications}, 2015, pp. 510--528.

\bibitem{schulman2017proximal}
J.~Schulman, F.~Wolski, P.~Dhariwal, A.~Radford, and O.~Klimov, ``Proximal
  policy optimization algorithms,'' \emph{arXiv preprint arXiv:1707.06347},
  2017.

\bibitem{shao2024deepseekmath}
Z.~Shao, P.~Wang, Q.~Zhu, R.~Xu, J.~Song, X.~Bi, H.~Zhang, M.~Zhang, Y.~Li,
  Y.~Wu \emph{et~al.}, ``Deepseekmath: Pushing the limits of mathematical
  reasoning in open language models,'' \emph{arXiv preprint arXiv:2402.03300},
  2024.

\bibitem{yao2025arctic}
Z.~Yao, G.~Sun, L.~Borchmann, Z.~Shen, M.~Deng, B.~Zhai, H.~Zhang, A.~Li, and
  Y.~He, ``Arctic-text2sql-r1: Simple rewards, strong reasoning in
  text-to-sql,'' \emph{arXiv preprint arXiv:2505.20315}, 2025.

\bibitem{zhai2025excot}
B.~Zhai, C.~Xu, Y.~He, and Z.~Yao, ``Excot: Optimizing reasoning for
  text-to-sql with execution feedback,'' \emph{arXiv preprint
  arXiv:2503.19988}, 2025.

\bibitem{nguyen2022model}
T.~T. Nguyen, T.~C. Phan, M.~H. Nguyen, M.~Weidlich, H.~Yin, J.~Jo, and
  Q.~V.~H. Nguyen, ``Model-agnostic and diverse explanations for streaming
  rumour graphs,'' \emph{Knowledge-Based Systems}, vol. 253, p. 109438, 2022.

\bibitem{duong2022deep}
C.~T. Duong, T.~T. Nguyen, T.-D. Hoang, H.~Yin, M.~Weidlich, and Q.~V.~H.
  Nguyen, ``Deep mincut: Learning node embeddings from detecting communities,''
  \emph{Pattern Recognition}, p. 109126, 2022.

\bibitem{finersql}
T.~D. Hoang, T.~T. Huynh, M.~Weidlich, T.~T. Nguyen, T.~Chen, H.~Yin, and
  Q.~V.~H. Nguyen, ``Boosting small language models for text-to-sql with
  fine-grained execution feedback and cost-efficient rewards,'' in
  \emph{ICDE}.\hskip 1em plus 0.5em minus 0.4em\relax IEEE, 2026.

\bibitem{guo2025deepseek}
D.~Guo, D.~Yang, H.~Zhang, J.~Song, R.~Zhang, R.~Xu, Q.~Zhu, S.~Ma, P.~Wang,
  X.~Bi \emph{et~al.}, ``Deepseek-r1: Incentivizing reasoning capability in
  llms via reinforcement learning,'' \emph{arXiv preprint arXiv:2501.12948},
  2025.

\bibitem{bai2022training}
Y.~Bai, A.~Jones, K.~Ndousse, A.~Askell, A.~Chen, N.~DasSarma, D.~Drain,
  S.~Fort, D.~Ganguli, T.~Henighan \emph{et~al.}, ``Training a helpful and
  harmless assistant with reinforcement learning from human feedback,''
  \emph{arXiv preprint arXiv:2204.05862}, 2022.

\bibitem{nguyen2014reconciling}
Q.~V.~H. Nguyen, T.~Nguyen~Thanh, Z.~Mikl{\'o}s, and K.~Aberer, ``Reconciling
  schema matching networks through crowdsourcing,'' \emph{EAI Endorsed
  Transactions on Collaborative Computing}, vol.~1, no.~2, p.~e2, 2014.

\bibitem{nguyen2023isomorphic}
T.~T. Nguyen, T.~T. Nguyen, T.~H. Nguyen, H.~Yin, T.~T. Nguyen, J.~Jo, and
  Q.~V.~H. Nguyen, ``Isomorphic graph embedding for progressive maximal
  frequent subgraph mining,'' \emph{ACM Transactions on Intelligent Systems and
  Technology}, vol.~15, no.~1, pp. 1--26, 2023.

\bibitem{rafailov2023direct}
R.~Rafailov, A.~Sharma, E.~Mitchell, C.~D. Manning, S.~Ermon, and C.~Finn,
  ``Direct preference optimization: Your language model is secretly a reward
  model,'' \emph{NeurIPS}, vol.~36, pp. 53\,728--53\,741, 2023.

\bibitem{hong2024orpo}
J.~Hong, N.~Lee, and J.~Thorne, ``Orpo: Monolithic preference optimization
  without reference model,'' in \emph{EMNLP}, 2024, pp. 11\,170--11\,189.

\bibitem{huynh2024fast}
T.~T. Huynh, T.~B. Nguyen, P.~L. Nguyen, T.~T. Nguyen, M.~Weidlich, Q.~V.~H.
  Nguyen, and K.~Aberer, ``Fast-fedul: A training-free federated unlearning
  with provable skew resilience,'' in \emph{Joint European Conference on
  Machine Learning and Knowledge Discovery in Databases}.\hskip 1em plus 0.5em
  minus 0.4em\relax Springer, 2024, pp. 55--72.

\bibitem{zhao2021eires}
B.~Zhao, H.~van~der Aa, T.~T. Nguyen, Q.~V.~H. Nguyen, and M.~Weidlich,
  ``Eires: Efficient integration of remote data in event stream processing,''
  in \emph{Proceedings of the 2021 International Conference on Management of
  Data}, 2021, pp. 2128--2141.

\bibitem{zhong2017seq2sql}
V.~Zhong, C.~Xiong, and R.~Socher, ``Seq2sql: Generating structured queries
  from natural language using reinforcement learning,'' \emph{arXiv preprint
  arXiv:1709.00103}, 2017.

\bibitem{xu2017sqlnet}
X.~Xu, C.~Liu, and D.~Song, ``Sqlnet: Generating structured queries from
  natural language without reinforcement learning,'' \emph{arXiv preprint
  arXiv:1711.04436}, 2017.

\bibitem{guo2019towards}
J.~Guo, Z.~Zhan, Y.~Gao, Y.~Xiao, J.-G. Lou, T.~Liu, and D.~Zhang, ``Towards
  complex text-to-sql in cross-domain database with intermediate
  representation,'' in \emph{Proceedings of the 57th Annual Meeting of the
  Association for Computational Linguistics}, 2019, pp. 4524--4535.

\bibitem{yu2018syntaxsqlnet}
T.~Yu, M.~Yasunaga, K.~Yang, R.~Zhang, D.~Wang, Z.~Li, and D.~Radev,
  ``{S}yntax{SQLN}et: Syntax tree networks for complex and cross-domain
  text-to-{SQL} task,'' in \emph{EMNLP}, 2018, pp. 1653--1663.

\bibitem{scholak2021picard}
T.~Scholak, N.~Schucher, and D.~Bahdanau, ``Picard: Parsing incrementally for
  constrained auto-regressive decoding from language models,'' in \emph{EMNLP},
  2021, pp. 9895--9901.

\bibitem{thang2015evaluation}
D.~C. Thang, N.~T. Tam, N.~Q.~V. Hung, and K.~Aberer, ``An evaluation of
  diversification techniques,'' in \emph{International Conference on Database
  and Expert Systems Applications}, 2015, pp. 215--231.

\bibitem{nguyen2020factcatch}
T.~T. Nguyen, M.~Weidlich, H.~Yin, B.~Zheng, Q.~H. Nguyen, and Q.~V.~H. Nguyen,
  ``Factcatch: Incremental pay-as-you-go fact checking with minimal user
  effort,'' in \emph{Proceedings of the 43rd International ACM SIGIR Conference
  on Research and Development in Information Retrieval}, 2020, pp. 2165--2168.

\bibitem{sun2023sql}
R.~Sun, S.~{\"O}. Arik, H.~Nakhost, H.~Dai, R.~Sinha, P.~Yin, and T.~Pfister,
  ``Sql-palm: Improved large language model adaptation for text-to-sql,''
  \emph{CoRR}, 2023.

\bibitem{pourreza2023dinsql}
M.~Pourreza and D.~Rafiei, ``Din-sql: Decomposed in-context learning of
  text-to-sql with self-correction,'' 2023.

\bibitem{nguyen2025device}
M.~H. Nguyen, T.~T. Huynh, T.~T. Nguyen, P.~L. Nguyen, H.~T. Pham, J.~Jo, and
  T.~T. Nguyen, ``On-device diagnostic recommendation with heterogeneous
  federated blocknets,'' \emph{Science China Information Sciences}, vol.~68,
  no.~4, p. 140102, 2025.

\bibitem{pham2025multilingual}
K.~T. Pham, T.~H. Nguyen, J.~Jo, Q.~V.~H. Nguyen, and T.~T. Nguyen,
  ``Multilingual text-to-sql: Benchmarking the limits of language models with
  collaborative language agents,'' in \emph{Australasian Database
  Conference}.\hskip 1em plus 0.5em minus 0.4em\relax Springer, 2025, pp.
  108--123.

\bibitem{nguyen2024multi}
D.~D.~A. Nguyen, M.~H. Nguyen, P.~L. Nguyen, J.~Jo, H.~Yin, and T.~T. Nguyen,
  ``Multi-task learning of heterogeneous hypergraph representations in lbsns,''
  in \emph{International Conference on Advanced Data Mining and
  Applications}.\hskip 1em plus 0.5em minus 0.4em\relax Springer, 2024, pp.
  161--177.

\bibitem{lyu2025sql}
S.~Lyu, H.~Luo, R.~Li, Z.~Ou, J.~Sun, Y.~Qin, X.~Shang, M.~Song, and Y.~Zhu,
  ``Sql-o1: A self-reward heuristic dynamic search method for text-to-sql,''
  \emph{arXiv preprint arXiv:2502.11741}, 2025.

\bibitem{thang2022nature}
D.~C. Thang, H.~T. Dat, N.~T. Tam, J.~Jo, N.~Q.~V. Hung, and K.~Aberer,
  ``Nature vs. nurture: Feature vs. structure for graph neural networks,''
  \emph{PRL}, vol. 159, pp. 46--53, 2022.

\bibitem{trung2022learning}
H.~T. Trung, T.~Van~Vinh, N.~T. Tam, J.~Jo, H.~Yin, and N.~Q.~V. Hung,
  ``Learning holistic interactions in lbsns with high-order, dynamic, and
  multi-role contexts,'' \emph{IEEE Transactions on Knowledge and Data
  Engineering}, vol.~35, no.~5, pp. 5002--5016, 2022.

\bibitem{huynh2021network}
T.~T. Huynh, C.~T. Duong, T.~T. Nguyen, V.~T. Van, A.~Sattar, H.~Yin, and
  Q.~V.~H. Nguyen, ``Network alignment with holistic embeddings,'' \emph{TKDE},
  vol.~35, no.~2, pp. 1881--1894, 2021.

\bibitem{nguyen2018if}
Q.~V.~H. Nguyen, K.~Zheng, M.~Weidlich, B.~Zheng, H.~Yin, T.~T. Nguyen, and
  B.~Stantic, ``What-if analysis with conflicting goals: Recommending data
  ranges for exploration,'' in \emph{2018 IEEE 34th International Conference on
  Data Engineering (ICDE)}.\hskip 1em plus 0.5em minus 0.4em\relax IEEE, 2018,
  pp. 89--100.

\bibitem{toan2018diversifying}
N.~T. Toan, P.~T. Cong, N.~T. Tam, N.~Q.~V. Hung, and B.~Stantic,
  ``Diversifying group recommendation,'' \emph{IEEE Access}, vol.~6, pp.
  17\,776--17\,786, 2018.

\bibitem{hung2017answer}
N.~Q.~V. Hung, D.~C. Thang, N.~T. Tam, M.~Weidlich, K.~Aberer, H.~Yin, and
  X.~Zhou, ``Answer validation for generic crowdsourcing tasks with minimal
  efforts,'' \emph{The VLDB Journal}, vol.~26, pp. 855--880, 2017.

\bibitem{nguyen2017argument}
Q.~V.~H. Nguyen, C.~T. Duong, T.~T. Nguyen, M.~Weidlich, K.~Aberer, H.~Yin, and
  X.~Zhou, ``Argument discovery via crowdsourcing,'' \emph{The VLDB Journal},
  vol.~26, no.~4, pp. 511--535, 2017.

\bibitem{nguyen2023example}
T.~T. Nguyen, T.~C. Phan, H.~T. Pham, T.~T. Nguyen, J.~Jo, and Q.~V.~H. Nguyen,
  ``Example-based explanations for streaming fraud detection on graphs,''
  \emph{Information Sciences}, vol. 621, pp. 319--340, 2023.

\bibitem{wei2022chain}
J.~Wei, X.~Wang, D.~Schuurmans, M.~Bosma, F.~Xia, E.~Chi, Q.~V. Le, D.~Zhou
  \emph{et~al.}, ``Chain-of-thought prompting elicits reasoning in large
  language models,'' \emph{NeurIPS}, vol.~35, pp. 24\,824--24\,837, 2022.

\bibitem{li2024dawn}
B.~Li, Y.~Luo, C.~Chai, G.~Li, and N.~Tang, ``The dawn of natural language to
  sql: Are we fully ready?'' \emph{PVLDB}, vol.~17, no.~11, pp. 3318--3331,
  2024.

\bibitem{murthy2024evaluating}
R.~Murthy, P.~Kumar, P.~Venkateswaran, and D.~Contractor, ``Evaluating the
  instruction-following abilities of language models using knowledge tasks,''
  \emph{arXiv preprint arXiv:2410.12972}, 2024.

\bibitem{qin-etal-2024-infobench}
Y.~Qin, K.~Song, Y.~Hu, W.~Yao, S.~Cho, X.~Wang, X.~Wu, F.~Liu, P.~Liu, and
  D.~Yu, ``{I}n{F}o{B}ench: Evaluating instruction following ability in large
  language models,'' in \emph{ACL}, 2024, pp. 13\,025--13\,048.

\bibitem{zhou2023instruction}
J.~Zhou, T.~Lu, S.~Mishra, S.~Brahma, S.~Basu, Y.~Luan, D.~Zhou, and L.~Hou,
  ``Instruction-following evaluation for large language models,'' \emph{arXiv
  preprint arXiv:2311.07911}, 2023.

\bibitem{nguyen2025finetuning}
X.-B. Nguyen, X.-H. Phan, and M.~Piccardi, ``Fine-tuning text-to-sql models
  with reinforcement-learning training objectives,'' \emph{Natural Language
  Processing Journal}, vol.~10, p. 100135, 2025.

\bibitem{lightman2024letsverify}
H.~Lightman, V.~Kosaraju, Y.~Burda, H.~Edwards, B.~Baker, T.~Lee, J.~Leike,
  J.~Schulman, I.~Sutskever, and K.~Cobbe, ``Let's verify step by step,'' in
  \emph{ICLR}, 2024.

\bibitem{yuan2024self}
W.~Yuan, R.~Y. Pang, K.~Cho, X.~Li, S.~Sukhbaatar, J.~Xu, and J.~E. Weston,
  ``Self-rewarding language models,'' in \emph{ICML}, 2024.

\bibitem{lu2024autopsv}
J.~Lu, Z.~Dou, H.~Wang, Z.~Cao, J.~Dai, Y.~Feng, and Z.~Guo, ``Autopsv:
  Automated process-supervised verifier,'' \emph{NeurIPS}, vol.~37, pp.
  79\,935--79\,962, 2024.

\bibitem{wang2024math}
P.~Wang, L.~Li, Z.~Shao, R.~Xu, D.~Dai, Y.~Li, D.~Chen, Y.~Wu, and Z.~Sui,
  ``Math-shepherd: Verify and reinforce llms step-by-step without human
  annotations,'' in \emph{ACL}, 2024, pp. 9426--9439.

\bibitem{li2025alphasql}
B.~Li, J.~Zhang, J.~Fan, Y.~Xu, C.~Chen, N.~Tang, and Y.~Luo, ``Alpha-sql:
  Zero-shot text-to-sql using monte carlo tree search,'' in \emph{ICML}, 2025.

\bibitem{pourreza2025reasoning}
M.~Pourreza, S.~Talaei, R.~Sun, X.~Wan, H.~Li, A.~Mirhoseini, A.~Saberi,
  S.~Arik \emph{et~al.}, ``Reasoning-sql: Reinforcement learning with sql
  tailored partial rewards for reasoning-enhanced text-to-sql,'' \emph{arXiv
  preprint arXiv:2503.23157}, 2025.

\bibitem{hoang2025scaling}
T.~D. Hoang, T.~T. Nguyen, T.~T. Huynh, H.~Yin, and Q.~V.~H. Nguyen, ``Scaling
  text2sql via llm-efficient schema filtering with functional dependency graph
  rerankers,'' \emph{arXiv preprint arXiv:2512.16083}, 2025.

\bibitem{han2023comprehensive}
Y.~Han, C.~Liu, and P.~Wang, ``A comprehensive survey on vector database:
  Storage and retrieval technique, challenge,'' \emph{arXiv preprint
  arXiv:2310.11703}, 2023.

\bibitem{zhang2025qwen3}
Y.~Zhang, M.~Li, D.~Long, X.~Zhang, H.~Lin, B.~Yang, P.~Xie, A.~Yang, D.~Liu,
  J.~Lin \emph{et~al.}, ``Qwen3 embedding: Advancing text embedding and
  reranking through foundation models,'' \emph{arXiv preprint
  arXiv:2506.05176}, 2025.

\bibitem{johnson2019billion}
J.~Johnson, M.~Douze, and H.~J{\'e}gou, ``Billion-scale similarity search with
  gpus,'' \emph{T-BD}, vol.~7, no.~3, pp. 535--547, 2019.

\bibitem{liu2025nl2sql}
X.~Liu, S.~Shen, B.~Li, N.~Tang, and Y.~Luo, ``Nl2sql-bugs: A benchmark for
  detecting semantic errors in nl2sql translation,'' in \emph{Proceedings of
  the 31st ACM SIGKDD Conference on Knowledge Discovery and Data Mining V. 2},
  2025, pp. 5662--5673.

\bibitem{ren2024comprehensive}
Z.~Ren, Y.~Chang, T.~T. Nguyen, Y.~Tan, K.~Qian, and B.~W. Schuller, ``A
  comprehensive survey on heart sound analysis in the deep learning era,''
  \emph{IEEE Computational Intelligence Magazine}, vol.~19, no.~3, pp. 42--57,
  2024.

\bibitem{nguyen2026review}
T.~T. Nguyen, Z.~Ren, T.~Pham, P.~L. Nguyen, Q.~V.~H. Nguyen, and H.~Yin, ``A
  review of instruction-guided image editing,'' \emph{EAAI}, 2026.

\bibitem{pham2025extensible}
M.~T. Pham, Q.~V.~H. Nguyen, J.~Jo, and T.~T. Nguyen, ``An extensible benchmark
  for value ambiguity resolution in text-to-sql,'' in \emph{Australasian
  Database Conference}.\hskip 1em plus 0.5em minus 0.4em\relax Springer, 2025,
  pp. 124--138.

\bibitem{huynh2025certified}
T.~T. Huynh, T.~B. Nguyen, T.~T. Nguyen, P.~L. Nguyen, H.~Yin, Q.~V.~H. Nguyen,
  and T.~T. Nguyen, ``Certified unlearning for federated recommendation,''
  \emph{ACM Transactions on Information Systems}, 2025.

\bibitem{yang2024pdc}
C.~Yang, W.~Yuan, L.~Qu, and T.~T. Nguyen, ``Pdc-frs: Privacy-preserving data
  contribution for federated recommender system,'' in \emph{International
  Conference on Advanced Data Mining and Applications}.\hskip 1em plus 0.5em
  minus 0.4em\relax Springer, 2024, pp. 65--79.

\bibitem{sakong2024higher}
D.~Sakong, V.~H. Vu, T.~T. Huynh, P.~Le~Nguyen, H.~Yin, Q.~V.~H. Nguyen, and
  T.~T. Nguyen, ``Higher-order knowledge-enhanced recommendation with
  heterogeneous hypergraph multi-attention,'' \emph{Information Sciences}, vol.
  680, p. 121165, 2024.

\bibitem{li2025omnisql}
H.~Li, S.~Wu, X.~Zhang, X.~Huang, J.~Zhang, F.~Jiang, S.~Wang, T.~Zhang,
  J.~Chen, R.~Shi \emph{et~al.}, ``Omnisql: Synthesizing high-quality
  text-to-sql data at scale,'' \emph{arXiv preprint arXiv:2503.02240}, 2025.

\bibitem{gretel-synthetic-text-to-sql-2024}
\BIBentryALTinterwordspacing
Y.~Meyer, M.~Emadi, D.~Nathawani, L.~Ramaswamy, K.~Boyd, M.~Van~Segbroeck,
  M.~Grossman, P.~Mlocek, and D.~Newberry, ``{Synthetic-Text-To-SQL}: A
  synthetic dataset for training language models to generate sql queries from
  natural language prompts,'' April 2024. [Online]. Available:
  \url{https://huggingface.co/datasets/gretelai/synthetic-text-to-sql}
\BIBentrySTDinterwordspacing

\bibitem{pourreza2023din}
M.~Pourreza and D.~Rafiei, ``Din-sql: Decomposed in-context learning of
  text-to-sql with self-correction,'' \emph{NeurIPS}, vol.~36, pp.
  36\,339--36\,348, 2023.

\bibitem{gao2023text}
D.~Gao, H.~Wang, Y.~Li, X.~Sun, Y.~Qian, B.~Ding, and J.~Zhou, ``Text-to-sql
  empowered by large language models: A benchmark evaluation,'' \emph{arXiv
  preprint arXiv:2308.15363}, 2023.

\bibitem{wang2023mac}
B.~Wang, C.~Ren, J.~Yang, X.~Liang, J.~Bai, Q.-W. Zhang, Z.~Yan, and Z.~Li,
  ``Mac-sql: Multi-agent collaboration for text-to-sql,'' \emph{arXiv preprint
  arXiv:2312.11242}, 2023.

\bibitem{lee2025mcs}
D.~Lee, C.~Park, J.~Kim, and H.~Park, ``Mcs-sql: Leveraging multiple prompts
  and multiple-choice selection for text-to-sql generation,'' in \emph{COLING},
  2025, pp. 337--353.

\bibitem{ma2025sql}
P.~Ma, X.~Zhuang, C.~Xu, X.~Jiang, R.~Chen, and J.~Guo, ``Sql-r1: Training
  natural language to sql reasoning model by reinforcement learning,''
  \emph{arXiv preprint arXiv:2504.08600}, 2025.

\bibitem{dao2022flashattention}
T.~Dao, D.~Fu, S.~Ermon, A.~Rudra, and C.~R{\'e}, ``Flashattention: Fast and
  memory-efficient exact attention with io-awareness,'' \emph{NeurIPS},
  vol.~35, pp. 16\,344--16\,359, 2022.

\bibitem{kwon2023efficient}
W.~Kwon, Z.~Li, S.~Zhuang, Y.~Sheng, L.~Zheng, C.~H. Yu, J.~E. Gonzalez,
  H.~Zhang, and I.~Stoica, ``Efficient memory management for large language
  model serving with pagedattention,'' in \emph{SIGOPS}, 2023.

\bibitem{zhu2018texygen}
Y.~Zhu, S.~Lu, L.~Zheng, J.~Guo, W.~Zhang, J.~Wang, and Y.~Yu, ``Texygen: A
  benchmarking platform for text generation models,'' in \emph{The 41st
  international ACM SIGIR conference on research \& development in information
  retrieval}, 2018, pp. 1097--1100.

\end{thebibliography}

\end{document}